\documentclass[12pt]{article}
\usepackage[super,comma,sort&compress]{natbib}
\usepackage[left=3cm,top=5cm,right=3cm,bottom=5cm,head=.1in]{geometry}
\usepackage{amssymb}
\usepackage{amsmath}
\usepackage{MnSymbol}
\usepackage{graphicx}
\usepackage{float}
\usepackage{ragged2e}
\usepackage{setspace}
\usepackage{blindtext}
\usepackage{bbm}
\usepackage{latexsym}
\usepackage{caption}
\usepackage{hyperref}
\usepackage{float}
\usepackage{comment}
\usepackage{dcolumn}
\usepackage{adjustbox}
\usepackage{xcolor}
\usepackage{setspace}
\usepackage{authblk}
\usepackage{subcaption}
\usepackage{booktabs}
\usepackage{multirow}
\usepackage{url}
\newtheorem{theorem}{Theorem}[section]

\title{Randomization-based Inference for Bernoulli-Trial Experiments and Implications for Observational Studies}
\author{Zach Branson and Marie-Ab\`ele Bind\thanks{
    This research was supported by the National Science Foundation Graduate Research Fellowship Program under Grant No. 1144152, and by the Office of the Director, National Institutes of Health under Award Number DP5OD021412. Any opinions, findings, and conclusions or recommendations expressed in this material are those of the authors and do not necessarily reflect the views of the National Science Foundation or the National Institutes of Health.}}
\affil{Harvard University}

\date{}

\begin{document}

\maketitle

\noindent
We present a randomization-based inferential framework for experiments characterized by a strongly ignorable assignment mechanism where units have independent probabilities of receiving treatment. Previous works on randomization tests often assume these probabilities are equal within blocks of units. We consider the general case where they differ across units and show how to perform randomization tests and obtain point estimates and confidence intervals. Furthermore, we develop rejection-sampling and importance-sampling approaches for conducting randomization-based inference conditional on any statistic of interest, such as the number of treated units or forms of covariate balance. We establish that our randomization tests are valid tests, and through simulation we demonstrate how the rejection-sampling and importance-sampling approaches can yield powerful randomization tests and thus precise inference. Our work also has implications for observational studies, which commonly assume a strongly ignorable assignment mechanism. Most methodologies for observational studies make additional modeling or asymptotic assumptions, while our framework only assumes the strongly ignorable assignment mechanism, and thus can be considered a minimal-assumption approach.

\section{Introduction}

Randomization-based inference centers around the idea that the treatment assignment mechanism is the only stochastic element in a randomized experiment and thus acts as the basis for conducting statistical inference.\citep{fisher1935design} In general, a central tenet of randomization-based inference is that the analysis of any given experiment should reflect its design: The inference for completely randomized experiments, blocked randomized experiments, and other designs should reflect the actual assignment mechanism that was used during the experiment. The idea that the assignment mechanism is the only stochastic element of an experiment is also commonly employed in the potential outcomes framework,\citep{neyman1923} which is now regularly used when estimating causal effects in randomized experiments and observational studies.\citep{rubin1974estimating,rubin2005causal} While randomization-based inference focuses on estimating causal effects for only the finite sample at hand, it can flexibly incorporate any kind of assignment mechanism without model specifications. Rosenbaum\cite{rosenbaum2002observational} provides a comprehensive review of randomization-based inference.

An essential step to estimating causal effects within the randomization-based inference framework as well as the potential outcomes framework is to state the probability distribution of the assignment mechanism. For simplicity, we focus on treatment-versus-control experiments, but our discussion can be extended to experiments with multiple treatments. Let the vector $\mathbf{W}$ denote the assignment mechanism for $N$ units in an experiment or observational study. It is commonly assumed that the probability distribution of $\mathbf{W}$ can be written as a product of independent Bernoulli trials that may depend on background covariates:\citep{rosenbaum2002covariance,rubin2007design,rubin2008objective}
\begin{align}
    P(\mathbf{W} = \mathbf{w} | \mathbf{X}) = \prod_{i=1}^N e(\mathbf{x}_i)^{w_i} [1 - e(\mathbf{x}_i)]^{1 - w_i}, \hspace{0.05 in } \text{where } 0 < e(\mathbf{x}_i) < 1 \hspace{0.05 in} \forall i =1,\dots,N \label{eqn:psModel}
\end{align}
Here, $\mathbf{X}$ is a $N \times p$ covariate matrix with rows $\mathbf{x}_i$, and $e(\mathbf{x}_i)$ denotes the probability that the $i^{\text{th}}$ unit receives treatment conditional on pre-treatment covariates $\mathbf{x}_i$; i.e., $e(\mathbf{x}_i) \equiv P(W_i = 1 | \mathbf{x}_i)$. The probabilities $e(\mathbf{x}_i)$ are commonly known as propensity scores.\citep{rosenbaum1983central} An assignment mechanism that can be written as (\ref{eqn:psModel}) is known as an unconfounded, strongly ignorable assignment mechanism.\citep{rubin2008objective} The assumption of an unconfounded, strongly ignorable assignment mechanism is essential to propensity score analyses and other methodologies (e.g., regression-based methods) for analyzing observational studies.\citep{dehejia2002propensity,sekhon2009opiates,stuart2010matching,austin2011introduction}

In randomized experiments, the propensity scores are defined by the designer(s) of the experiment and are thus known; this knowledge is all that is needed to construct unbiased estimates for average treatment effects.\citep{rubin2008objective} The propensity score $e(\mathbf{x}_i)$ is not necessarily a function of all or any of the covariates: For example, in completely randomized experiments, $e(\mathbf{x}_i) = 0.5$ for all units; and for blocked-randomized and paired experiments, the propensity scores are equal for all units within the same block or pair.

In observational studies, the propensity scores are not known, and instead must be estimated. The $e(\mathbf{x}_i)$ in (\ref{eqn:psModel}) are often estimated using logistic regression, but any model that estimates conditional probabilities for a binary treatment can be used. These estimates, $\hat{e}(\mathbf{x}_i)$, are commonly employed to ``reconstruct'' a hypothetical experiment that yielded the observed data.\citep{rubin2008objective} For example, matching methodologies are used to obtain subsets of treatment and control that are balanced in terms of pre-treatment covariates; then, these subsets of treatment and control are analyzed as if they came from a completely randomized experiment.\citep{ho2007matching,rubin2008objective,stuart2010matching} Others have suggested regression-based adjustments combined with the propensity score\cite{robins1995semiparametric,rubin2000combining} as well as Bayesian modeling.\cite{rubin1978bayesian,rubin2005causal,zigler2014uncertainty} Notably, all of these methodologies implicitly assume the Bernoulli trial assignment mechanism shown in (\ref{eqn:psModel}), but the subsequent analyses reflect a completely randomized, blocked-randomized, or paired assignment mechanism instead. One methodology commonly employed in observational studies that more closely reflects a Bernoulli trial assignment mechanism is inverse propensity score weighting;\citep{hirano2001estimation,hirano2003efficient,lunceford2004stratification,hernan2006estimating} however, the variance of such estimators is unstable, especially when estimated propensity scores are particularly close to 0 or 1, which is an ongoing concern in the literature.\citep{cole2008constructing,austin2015moving} Furthermore, the validity of such point estimates and uncertainty intervals rely on asymptotic arguments and an infinite-population interpretation.

More importantly, all of the above methodologies---matching, frequentist or Bayesian modeling, inverse propensity score weighting, or any combination of them---assume the strongly ignorable assignment mechanism shown in (\ref{eqn:psModel}), but they also intrinsically make additional modeling or asymptotic assumptions. On the other hand, although randomization-based inference methodologies also make the common assumption of the strongly ignorable assignment mechanism, they do not require any additional model specifications or asymptotic arguments.

However, while there is a wide literature on randomization tests, most have focused on assignment mechanisms where the propensity scores are assumed to be the same across units (i.e., completely randomized experiments) or groups of units (i.e., blocked or paired experiments), instead of the more general case where they may differ across all units, as in (\ref{eqn:psModel}). Imbens and Rubin\cite{imbens2015causal} briefly mention Bernoulli trial experiments, but only discuss inference for purely randomized and block randomized designs. Another example is Basu,\cite{basu1980randomization} who thoroughly discusses Fisherian randomization tests and briefly considers Bernoulli trial experiments, but does not provide a randomization-test framework for such experiments. This trend continues for observational studies: Most randomization tests for observational studies utilize permutations of the treatment indicator within covariate strata, and thus reflect a block-randomized assignment mechanism instead of the assumed Bernoull trial assignment mechanism.\citep{rosenbaum1984conditional,rosenbaum1988permutation,rosenbaum2002covariance} While these tests are valid under certain assumptions, they are not immediately applicable to cases where covariates are not easily stratified (e.g., continuous covariates) or where there is not at least one treated unit and one control unit in each stratum.\cite{rosenbaum2002observational} None of these randomization tests are applicable to cases where the propensity scores (known or unknown) differ across all units.

Most randomization tests that incorporate varying propensity scores focus on the biased-coin design popularized by Efron\cite{efron1971forcing}, where propensity scores are dependent on the order units enter the experiment and possibly pre-treatment covariates as well. Wei\cite{wei1978application} and Soares and Wu\cite{soares1983some} developed extensions for this experimental design, while Smythe and Wei\cite{smythe1983significance}, Wei\cite{wei1988exact}, and Mehta et al.\cite{mehta1988constructing} developed significance tests for such designs. Good\cite{good2013permutation} (Section 4.5) provides further discussion on this literature. The biased-coin design is related to covariate-adaptive randomization schemes in the clinical trial literature, starting with the work of Pocock and Simon.\cite{pocock1975sequential} Covariate-adaptive randomization schemes sequentially randomize units such that the treatment and control groups are balanced in terms of pre-treatment covariates,\cite{loux2013simple,lin2015pursuit,zagoraiou2017choosing} and recent works in the statistics literature have explored valid randomization tests for covariate-adaptive randomization schemes.\cite{simon2011using,shao2013validity} Importantly, the randomization test literature for biased-coin and covariate-adaptive designs differs from the randomization test presented here: All of these works focus on sequential designs, and thus depend on the sequential dependence among units inherent in the randomization scheme. In contrast, we assume that all units are simultaneously assigned to treatment according to the strongly ignorable assignment mechanism (\ref{eqn:psModel}).

To the best of our knowledge, there is not an explicit randomization-based inference framework for analyzing Bernoulli trial experiments, let alone observational studies. Here we develop such a framework for randomized experiments characterized by Bernoulli trials, with the implication that this framework can be extended to the observational study literature as well. In particular, we develop rejection-sampling and importance-sampling approaches for conducting conditional randomization-based inference for Bernoull trial experiments, which has not been previously discussed in the literature. These approaches allow one to conduct randomization tests conditional on statistics of interest for more precise inference.

In Section \ref{s:randomizationInferenceReview}, we review randomization-based inference in general, including randomization tests and how these tests can be inverted to yield point estimates and confidence intervals. In Section \ref{s:bernoulliTrials}, we develop a randomization-based inference framework for Bernoulli trial experiments, first reviewing the case where propensity scores are equal across units, and then extending this framework to the general case where propensity scores differ across units. Furthermore, we establish that randomization tests under this framework are valid tests, both unconditionally and conditional on statistics of interest. In Section \ref{s:simulationExample}, we demonstrate our framework with a simple example and provide simulation evidence for how our rejection-sampling and importance-sampling approaches can yield statistically powerful conditional randomization tests. In Section \ref{s:discussion}, we discuss extensions and implications of this work, particularly for observational studies.

\section{Review of Randomization-Based Inference} \label{s:randomizationInferenceReview}

Randomization-based inference focuses on randomization tests for treatment effects, which can be inverted to obtain both point estimates and confidence intervals. Randomization tests were first proposed by Fisher,\cite{fisher1935design} and foundational theory for these tests was later developed by Pitman\cite{pitman1938significance} and Kempthorne.\cite{kempthorne1952design} We follow the notation of Imbens and Rubin\cite{imbens2015causal} in our discussion of randomization tests for treatment-versus-control experiments.

\subsection{Notation}

Randomization tests utilize the potential outcomes framework, where the only stochastic element of an experiment is the treatment assignment. Let
\begin{align}
    W_i = \begin{cases}
    1 &\mbox{ if } \text{the $i^{\text{th}}$ unit receives treatment}  \\
    0 &\mbox{ if } \text{the $i^{\text{th}}$ unit receives control}
    \end{cases}
\end{align}
denote the treatment assignment, and let $Y_i(W_i)$ denote the $i^{\text{th}}$ unit's potential outcome, which only depends on the treatment assignment $W_i$. Only $Y_i(1)$ or $Y_i(0)$ is ultimately observed at the end of an experiment---never both. Let
\begin{align}
    y_i^{obs} = Y_i(1) W_i + Y_i(0)(1 - W_i)
\end{align}
denote the observed outcomes. Finally, let $\mathbb{W} \equiv \{0, 1\}^N$ denote the set of all possible treatment assignments, and let $\mathbb{W}^+ \subset \mathbb{W}$ denote the subset of $\mathbb{W}$ with positive probability, i.e., $\mathbb{W}^+ = \{\mathbf{w} \in \mathbb{W} : P(\mathbf{W} = \mathbf{w}) > 0\}$.

Importantly, the probability distribution of treatment assignments, $P(\mathbf{W})$, fully characterizes the assignment mechanism: Because treatment assignment is the only stochastic element in a randomized experiment, the distribution $P(\mathbf{W})$ specifies the randomness in a randomized experiment. Consequentially, inference within the randomization-based framework is determined by $P(\mathbf{W})$.

We first review how $P(\mathbf{W})$ is used to perform randomization tests. We then discuss how to invert these tests to obtain point estimates and confidence intervals for the average treatment effect.

\subsection{Testing the Sharp Null Hypothesis via Randomization Tests} \label{ss:testingFishersSharpNull}

The most common use of randomization tests is to test the Sharp Null Hypothesis, which is
\begin{align}
    H_0: Y_i(1) = Y_i(0) \hspace{0.05 in} \forall i = 1, \dots, n \label{sharpNull}
\end{align}
i.e., the hypothesis that there is no treatment effect. Under the Sharp Null Hypothesis, the outcomes for \textit{any} randomization from the set of all possible randomizations $\mathbb{W}^+$ is known: Regardless of a unit's treatment assignment, its outcome will always be equal to the observed response $y_i^{obs}$ under the Sharp Null Hypothesis. This knowledge allows one to test the Sharp Null Hypothesis.

To test this hypothesis, one first chooses a suitable test statistic
\begin{align}
    t \big(Y(\mathbf{W}), \mathbf{W} \big) \label{testStatistic}
\end{align}
and determines whether the observed test statistic $t^{obs} \equiv t(\mathbf{y}^{obs}, \mathbf{W}^{obs})$ is unlikely to occur according to the randomization distribution of the test statistic (\ref{testStatistic}) under the Sharp Null Hypothesis. For example, one common choice of test statistic is the difference in mean response between treatment and control units, defined as
\begin{align}
    t \big(Y(\mathbf{W}), \mathbf{W} \big) = \frac{\sum_{i: W_i = 1} Y_i(1)}{\sum_{i=1}^N W_i} - \frac{\sum_{i: W_i = 0} Y_i(0)}{\sum_{i=1}^N (1-W_i)} \label{eqn:meanDiffEstimator}
\end{align}
Such a test statistic will be powerful in detecting a difference in means between the distributions of $Y_i(1)$ and $Y_i(0)$. In general, one should choose a test statistic according to possible differences in the distributions of $Y_i(1)$ and $Y_i(0)$ that one is most interested in. Please see Rosenbaum\cite{rosenbaum2002observational} (Chapter 2) for a discussion on the choice of test statistics for randomization tests.

After a test statistic is chosen, a randomization-test $p$-value can be computed by comparing the observed test statistic $t^{obs}$ to the set of $t \big( Y(\mathbf{W}), \mathbf{W} \big)$ that are possible given the set of possible treatment assignments $\mathbb{W}^+$, assuming the Sharp Null Hypothesis is true. The two-sided randomization-test $p$-value is
\begin{align}
    P \big( |t \big(Y(\mathbf{W}), \mathbf{W} \big)| \geq |t^{obs}| \big) &= \sum_{\mathbf{w} \in \mathbb{W}^+} \mathbb{I} \big( \big| t \big(Y(\mathbf{w}), \mathbf{w} \big) \big| \geq | t^{obs} | \big)P(\mathbf{W} = \mathbf{w}) \label{randomizationTestPValue}
\end{align}
where $\mathbb{I}(A) = 1$ if event $A$ occurs and zero otherwise. Importantly, the randomization-test $p$-value (\ref{randomizationTestPValue}) depends on the set of possible treatment assignments $\mathbb{W}^+$, the probability distribution $P(\mathbf{W})$, and the choice of test statistic $t \big( Y(\mathbf{W}), \mathbf{W} \big)$.

Thus, testing the Sharp Null Hypothesis is a three-step procedure:
\begin{enumerate}
    \item Specify the distribution $P(\mathbf{W})$ (and, consequentially, the set of possible treatment assignments $\mathbb{W}^+$).
    \item Choose a test statistic $t\big(Y(\mathbf{W}), \mathbf{W} \big)$.
    \item Compute or approximate the $p$-value (\ref{randomizationTestPValue}).
\end{enumerate}
All randomization tests discussed in this paper follow this three-step procedure, with the only difference among them being the choice of $P(\mathbf{W})$, i.e. the first step. The third step notes that exactly computing the randomization-test $p$-value is often computationally intensive because it requires enumerating all possible $\mathbf{W} \in \mathbb{W}^+$; instead, it can be approximated. A typical approximation is to generate a random sample $\mathbf{w}^{(1)}, \dots, \mathbf{w}^{(M)}$ from $P(\mathbf{W})$, and then approximate the $p$-value (\ref{randomizationTestPValue}) by
\begin{align}
    P \big( \big| t \big(Y(\mathbf{W}), \mathbf{W} \big) \big| \geq |t^{obs}| \big) &\approx \frac{ \sum_{m=1}^M \mathbb{I} \big( \big| t \big(Y(\mathbf{w}^{(m)}), \mathbf{w}^{(m)} \big) \big| \geq |t^{obs}| \big)}{M} \label{randomizationTestPValueApproximationSimple}
\end{align}
Importantly, the approximation (\ref{randomizationTestPValueApproximationSimple}) still depends on the probability distribution of the assignment mechanism, $P(\mathbf{W})$, because the random samples $\mathbf{w}^{(1)}, \dots, \mathbf{w}^{(M)}$ are generated using $P(\mathbf{W})$. This distinction will be important in our discussion of Bernoulli trial experiments, where the probability of receiving treatment---i.e., the propensity scores---may be equal or non-equal across units. In both cases, the set $\mathbb{W}^+$ is the same, but the probability distribution $P(\mathbf{W})$ is different.

Testing the Sharp Null Hypothesis will provide information about the presence of any treatment effect amongst all units in the study. Furthermore, this test can be inverted to obtain point estimates and confidence intervals for the treatment effect. 

\subsection{Randomization-based Point Estimates and Confidence Intervals for the Treatment Effect} \label{ss:confidenceIntervals}

A confidence interval can be constructed by inverting a variation of the Sharp Null Hypothesis that assumes an additive treatment effect. A randomization-based confidence interval for the average treatment effect is the set of $\tau \in \mathbb{R}$ such that one fails to reject the hypothesis
\begin{align}
    H_0^{\tau}: Y_i(1) = Y_i(0) + \tau \hspace{0.05 in} \forall i = 1, \dots, N \label{sharpNullTau}
\end{align}
The above hypothesis is a sharp hypothesis in the sense that, under $H_0^{\tau}$, every unit's outcome for any treatment assignment is known: Under $H_0^{\tau}$, the missing potential outcome of any treated unit would be $y_i^{obs} - \tau$; likewise, the missing potential outcome of any control unit would be $y_i^{obs} + \tau$. Thus, for any hypothetical treatment assignment $\mathbf{w} \in \mathbb{W}^+$, one can calculate the corresponding potential outcomes $Y(\mathbf{w})$ under $H_0^{\tau}$ in terms of the observed outcomes $\mathbf{y}^{obs}$ and observed treatment assignment $\mathbf{w}^{obs}$:
\begin{align}
	Y_i(w_i) &= y_i^{obs} + \tau (w_i - w_i^{obs}), \hspace{0.05 in} \forall i = 1, \dots, N \label{eqn:hypotheticalPotentialOutcomesTau}
\end{align}
Therefore, one can obtain a $p$-value for the hypothesis $H_0^{\tau}$ by drawing many hypothetical randomizations $\mathbf{w}^{(1)},\dots, \mathbf{w}^{(M)}$ from $P(\mathbf{W})$, computing each $Y(\mathbf{w}^{(m)})$ using (\ref{eqn:hypotheticalPotentialOutcomesTau}), and then using (\ref{randomizationTestPValueApproximationSimple}) to approximate the $p$-value for any given test statistic $t(Y(\mathbf{W}), \mathbf{W})$.

To construct a 95\% confidence interval, one considers many $\tau$ (e.g., via a line search), tests the hypothesis $H_0^{\tau}$ for each $\tau$, and defines the confidence interval as the set of $\tau$ with corresponding $p$-values above 0.05.\citep{rosenbaum2002observational,imbens2015causal} Importantly, the confidence interval will depend on the probability distribution $P(\mathbf{W})$ through the draws $\mathbf{w}^{(1)},\dots,\mathbf{w}^{(M)}$ to compute each $p$-value; thus, the confidence interval will reflect a prespecified assignment mechanism. As we discuss in Section \ref{ss:acceptRejectProcedure}, this also allows one to flexibly construct confidence intervals that condition on particular statistics of interest.

Testing the hypothesis $H_0^{\tau}$ also yields a natural point estimate: Define the point estimate $\hat{\tau}$ as the $\tau$ such that the $p$-value for testing the hypothesis $H_0^{\tau}$ is maximized. For example, given a 95\% confidence interval containing $\tau$ with corresponding $p$-values above 0.05, $\hat{\tau}$ is defined as the $\tau$ with the highest $p$-value. The interpretation of such a $\hat{\tau}$ is that this is the ``most probable'' $\tau$ under the assumption of an additive treatment effect. This point estimate is a variant of the Hodges-Lehmann randomization-based point estimate, which equates the test statistic under the hypothesis $H_0^{\tau}$ to its expectation under the randomization distribution.\citep{hodges1963estimates,rosenbaum2002observational}

Some have criticized randomization-based confidence intervals constructed by inverting hypotheses such as (\ref{sharpNullTau}) because it assumes a homogeneous treatment effect, which may be an inappropriate assumption. However, in general, confidence intervals can be constructed using any Sharp Null Hypothesis that fully specifies unit-level treatment effects, including sharp null hypotheses that specify heterogeneous treatment effects.\citep{caughey2016beyond} Thus, while we focus on homogeneous treatment effects as assumed in (\ref{sharpNullTau}), the randomization test framework that we present below can be extended to point estimates and confidence intervals that account for treatment effect heterogeneity to the extent that one can specify sharp null hypotheses that incorporate heterogeneous treatment effects.

\section{Randomization-based Inference for Bernoulli Trial Experiments} \label{s:bernoulliTrials}

Here we consider experimental designs that are characterized by Bernoulli trials and develop randomization tests for these designs. First, we review randomization tests for experimental designs where the probability of receiving treatment is the same for all units; this will motivate our development of randomization tests for experimental designs where the probability of receiving treatment differs across units, which is our main contribution. For both cases---first when the propensity scores are equal across units, and then when the propensity scores differ---we will discuss several assignment mechanisms $P(\mathbf{W})$ and sets of possible treatment assignments $\mathbb{W}^+$, which correspond to different randomization tests. Once $P(\mathbf{W})$ and $\mathbb{W}^+$ are specified, the Sharp Null Hypothesis can be tested by following the three-step procedure in Section \ref{ss:testingFishersSharpNull}; furthermore, these tests can be inverted to yield point estimates and confidence intervals, as discussed in Section \ref{ss:confidenceIntervals}. For each test, we will state an explicit form for $P(\mathbf{W} = \mathbf{w})$ for any $\mathbf{w} \in \mathbb{W}^+$ to compute the randomization test $p$-value (\ref{randomizationTestPValue}) exactly, and we will also state how random samples $\mathbf{w}^{(1)}, \dots, \mathbf{w}^{(M)}$ can be generated to approximate this $p$-value using (\ref{randomizationTestPValueApproximationSimple}). In Section \ref{ss:acceptRejectProcedure}, we introduce rejection-sampling and importance-sampling approaches to perform randomization tests conditional on various statistics of interest, which has not been previously considered for randomization-based inference for Bernoulli trial experiments.

\subsection{Case 1: Propensity Scores are Equal Across Units} \label{ss:equalProbabilities}

Let $e(\mathbf{x}_i) = P(\mathbf{W}_i = 1 | \mathbf{x}_i)$ denote the propensity score, i.e., the probability that the $i^{\text{th}}$ unit receives treatment, given a vector of pre-treatment covariates $\mathbf{x}_i$. In this section we assume without loss of generality that $e(\mathbf{x}_i) = 0.5$ for all $i = 1,\dots,N$; i.e., $P(W_i = 1 | \mathbf{x}_i) = P(W_i = 1) = 0.5$ for all units. We consider several sets of possible treatment assignments $\mathbb{W}^+$ and note the corresponding $P(\mathbf{W} = \mathbf{w})$ for each $\mathbf{w} \in \mathbb{W}^+$, which can be used to compute the $p$-value (\ref{randomizationTestPValue}) for testing the Sharp Null Hypothesis.

First consider the set $\mathbb{W}^+ = \mathbb{W} = \{0,1\}^N$, i.e., experiments that are characterized by independent, unbiased coin flips, where any number of units can receive treatment or control. In this case, $P(\mathbf{W} = \mathbf{w}) = \frac{1}{2^N}$ for all $\mathbf{w} \in \mathbb{W}^+$. To generate random draws $\mathbf{w}^{(1)}, \dots, \mathbf{w}^{(M)}$, one simply flips $N$ unbiased coins to generate an $N$-dimensional vector of 0s and 1s.

However, Imbens and Rubin\cite{imbens2015causal} note that when $\mathbb{W}^+ = \{0, 1\}^N$, there is a non-zero probability of $\mathbf{W} = \mathbf{0}_N \equiv (0, \dots, 0)$ or $\mathbf{W} = \mathbf{1}_N \equiv (1, \dots, 1)$. In these cases, most test statistics are undefined, and so they do not consider this case further. This concern can be addressed by either defining test statistics for these cases (a common choice being zero) or instead considering the set $\mathbb{W}^+ = \{0, 1\}^N \setminus (\mathbf{0}_N \cup \mathbf{1}_N )$ of possible treatment assignments. In this case, $P(\mathbf{W} = \mathbf{w}) = \frac{1}{2^N - 2}$ for all $\mathbf{w} \in \mathbb{W}^+$. To generate random draws $\mathbf{w}^{(1)}, \dots, \mathbf{w}^{(M)}$, one simply flips $N$ unbiased coins and only accepts a random draw $\mathbf{w}^{(m)}$ if it is not $\mathbf{0}_N$ or $\mathbf{1}_N$. This follows the argument of Imbens and Rubin\cite{imbens2015causal} that preventing ``unhelpful treatment allocations'' will yield more precise inferences for treatment effects.

Indeed, we can even further restrict $\mathbb{W}^+$. It is common to condition on statistics such as the number of units that receive treatment $N_T \equiv \sum_{i=1}^N W_i$. When $\mathbb{W}^+ = \{ \mathbf{W} \in \mathbb{W} | \sum_{i=1}^N W_i = N_T\}$ for some prespecified $N_T$, $P(\mathbf{W} = \mathbf{w}) = \frac{1}{ {N \choose N_T} }$ for all $\mathbf{w} \in \mathbb{W}^+$. To generate random draws $\mathbf{w}^{(1)}, \dots, \mathbf{w}^{(M)}$, one simply flips $N$ unbiased coins and only accepts a random draw $\mathbf{w}^{(m)}$ if $\sum_{i=1}^N w_i^{(m)} = N_T$; equivalently, one can obtain such random draws by randomly permuting the observed treatment assignment $\mathbf{W}^{obs}$. A randomization test that uses such a $\mathbb{W}^+$ and $P(\mathbf{W})$ is the most common randomization test in the literature and corresponds to what is typically referred to as a ``completely randomized'' experimental design.\citep{imbens2015causal} Because of the equivalence to random permutations of $\mathbf{W}^{obs}$, this randomization test is also often called a permutation test.

\subsection{Case 2: Propensity Scores Differ Across Units} \label{ss:unequalProbabilities}

Now consider the case where $e(\mathbf{x}_i) \neq e(\mathbf{x}_j)$ for some $i \neq j$, i.e., where the propensity scores differ across units. This may be due to differences in the covariate vectors $\mathbf{x}_i$ and $\mathbf{x}_j$ or some other experimental design prespecification. Again we consider several sets of possible treatment assignments $\mathbb{W}^+$, note the corresponding $P(\mathbf{W} = \mathbf{w} | \mathbf{X})$ for each $\mathbf{w} \in \mathbb{W}^+$, and state how to generate random draws $\mathbf{w}^{(1)}, \dots, \mathbf{w}^{(M)}$, which can be used to compute or approximate the $p$-value for testing the Sharp Null Hypothesis.

First consider the set $\mathbb{W}^+ = \mathbb{W} = \{0,1\}^N$. In this case,
\begin{align}
     P(\mathbf{W} = \mathbf{w} | \mathbf{X}) = \prod_{i=1}^N e(\mathbf{x}_i)^{w_i}[1 - e(\mathbf{x}_i)]^{1 - w_i}
 \end{align} 
which is identical to the assignment mechanism (\ref{eqn:psModel}) typically assumed in observational studies. To generate random draws $\mathbf{w}^{(1)}, \dots, \mathbf{w}^{(M)}$, one simply flips $N$ \textit{biased} coins with probabilities corresponding to the $e(\mathbf{x}_i)$ to generate an $N$-dimensional vector of 0s and 1s.

However, there is still a chance---though small---that a random draw $\mathbf{w}$ from $\mathbb{W}^+ = \{0, 1\}^N$ will be equal to $\mathbf{0}_N$ or $\mathbf{1}_N$, and in this case test statistics will be undefined. Now consider the restricted set $\mathbb{W}^+ = \{0, 1\}^N \setminus (\mathbf{0}_N \cup \mathbf{1}_N )$. In this case,
\begin{align}
     P(\mathbf{W} = \mathbf{w} | \mathbf{X}) = \frac{\prod_{i=1}^N e(\mathbf{x}_i)^{w_i}[1 - e(\mathbf{x}_i)]^{1 - w_i}}{1 - \prod_{i=1}^N e(\mathbf{x}_i) - \prod_{i=1}^N [1 - e(\mathbf{x}_i)]} \label{eqn:biasedCoinRestricted01Probabilities}
 \end{align}
To arrive at this result, note that when $\mathbb{W}^+ = \{0, 1\}^N \setminus (\mathbf{0}_N \cup \mathbf{1}_N )$,
\begin{align}
    \sum_{\mathbf{w} \in \mathbb{W}^+} \prod_{i=1}^N e(\mathbf{x}_i)^{w_i}[1 - e(\mathbf{x}_i)]^{1 - w_i} = 1 - \prod_{i=1}^N e(\mathbf{x}_i) - \prod_{i=1}^N [1 - e(\mathbf{x}_i)]
\end{align}
Thus, the probabilities (\ref{eqn:biasedCoinRestricted01Probabilities}) sum to one. To generate random draws $\mathbf{w}^{(1)}, \dots, \mathbf{w}^{(M)}$, one simply flips $N$ \textit{biased} coins and only accepts a random draw $\mathbf{w}^{(m)}$ if it is not $\mathbf{0}_N$ or $\mathbf{1}_N$.

Again, we can further restrict $\mathbb{W}^+$ to incorporate certain statistics of interest, such as the number of units assigned to treatment. Consider the set $\mathbb{W}^+ = \{\mathbf{W} \in \mathbb{W} | \sum_{i=1}^N W_i = N_T\}$ for some prespecified $N_T$. In this case,
\begin{align}
    P(\mathbf{W} = \mathbf{w} | \mathbf{X}) = \frac{\prod_{i=1}^N e(\mathbf{x}_i)^{w_i}[1 - e(\mathbf{x}_i)]^{1 - w_i}}{P(\sum_{i=1}^N W_i = N_T | \mathbf{X}) } \label{eqn:probabilityConditionalNt}
\end{align}
The denominator, $P(\sum_{i=1}^N W_i = N_T | \mathbf{X}) = \sum_{\mathbf{w} \in \mathbb{W}^+} \prod_{i=1}^N e(\mathbf{x}_i)^{w_i} [1 - e(\mathbf{x}_i)]^{1-w_i}$, is seemingly difficult to compute, due to the large number, ${N \choose N_T }$, of possible treatment assignments $\mathbf{w} \in \mathbb{W}^+$. Chen and Liu\cite{chen1997statistical} provide an algorithm to compute $P(\sum_{i=1}^N W_i = N_T | \mathbf{X})$ exactly. Alternatively, $P(\sum_{i=1}^N W_i = N_T | \mathbf{X})$ can be estimated, and there are many ways to estimate this quantity. One option is to randomly sample $\mathbf{w}^{(1)}, \dots, \mathbf{w}^{(M)}$ from $\mathbb{W}^+$ and use the unbiased estimator
\begin{align}
    \widehat{P} \left(\sum_{i=1}^N W_i = N_T | \mathbf{X} \right) = \frac{ {N \choose N_T} }{M} \sum_{m = 1}^M \prod_{i=1}^N e(\mathbf{x}_i)^{w_i^{(m)}} [1 - e(\mathbf{x}_i)]^{1-w_i^{(m)}} \label{sumApproximation}
\end{align}
which is the typical estimator for a population total seen in the survey sampling literature (e.g., Lohr Page 55).\cite{lohr2009sampling}

However, computing $P\left(\sum_{i=1}^N W_i = N_T | \mathbf{X} \right)$ is only required when one wants to compute the randomization-test $p$-value exactly using (\ref{randomizationTestPValue}). Instead, one can still approximate this $p$-value using (\ref{randomizationTestPValueApproximationSimple}) by generating random draws $\mathbf{w}^{(1)}, \dots, \mathbf{w}^{(M)}$, which is done by flipping $N$ \textit{biased} coins and only accepting a random draw $\mathbf{w}$ if $\sum_{i=1}^N w_i = N_T$.

This introduces straightforward rejection-sampling and importance-sampling procedures for conducting conditional randomization-based inference for Bernoulli trial experiments.

\subsection{Rejection-Sampling and Importance-Sampling Procedures for Conditional Randomization Tests} \label{ss:acceptRejectProcedure}

As discussed in Section \ref{s:randomizationInferenceReview}, researchers do not typically compute the randomization test $p$-value (\ref{randomizationTestPValue}) exactly, but instead generate random draws $\mathbf{w}^{(1)}, \dots, \mathbf{w}^{(M)}$ from the probability distribution $P(\mathbf{W})$ and then approximate the randomization test $p$-value using (\ref{randomizationTestPValueApproximationSimple}). To conduct conditional randomization-based inference, one generates random draws from conditional probability distributions such as $P(\mathbf{W} | \sum_{i=1}^N W_i = N_T)$ instead of $P(\mathbf{W})$. This is straightforward when the propensity scores are the same across units: For example, as discussed in Section \ref{ss:equalProbabilities}, samples from $P(\mathbf{W} | \sum_{i=1}^N W_i = N_T)$ correspond to random permutations of the observed treatment assignment $\mathbf{W}^{obs}$ when the propensity scores are equal across units. However, sampling from such conditional distributions when the propensity scores differ across units is less trivial. To the best of our knowledge, a strategy for how to sample from such distributions has not been described in the literature.

Conducting conditional randomization-based inference involves focusing only on ``acceptable'' treatment assignments $\mathbf{W}$; e.g., $\mathbf{W}$ that are not $\mathbf{0}_N$ or $\mathbf{1}_N$, or $\mathbf{W}$ such that $\sum_{i=1}^N W_i = N_T$ for some prespecified $N_T$. To formalize this idea, define an acceptance criterion that is a function of the treatment assignment and pre-treatment covariates:
\begin{align}
    \phi(\mathbf{W}, \mathbf{X}) = \begin{cases}
        1 &\mbox{ if } \mathbf{W} \text{ is an acceptable treatment assignment} \\
        0 &\mbox{ if } \mathbf{W} \text{ is not an acceptable treatment assignment.}
    \end{cases}
\end{align}
The criterion $\phi(\mathbf{W}, \mathbf{X})$ can encapsulate any statistic of interest, such as the number of treated units or forms of covariate balance. The criterion $\phi(\mathbf{W}, \mathbf{X})$ should be defined by statistics that are believed to be related to the outcome, such as the number of treated units with a certain covariate value or the covariate means in the treatment and control groups. See Hennessy et al.\cite{hennessy2016conditional} for further discussion about the types of statistics that should be conditioned on for conditional randomization-based inference.

Once $\phi(\mathbf{W}, \mathbf{X})$ is defined, one conducts conditional randomization-based inference by performing a randomization test only within the set of randomizations such that the acceptance criterion is satisfied. For example, Sections \ref{ss:equalProbabilities} and \ref{ss:unequalProbabilities} discuss conducting randomization-based inference for the case when $\phi(\mathbf{W}, \mathbf{X}) = 1$ if $\sum_{i=1}^N W_i = N_T$ and 0 otherwise. Thus, the true conditional randomization test $p$-value is
\begin{align}
    p_{\phi} \equiv \sum_{\mathbf{w} \in \mathbb{W}_{\phi}^+} \mathbb{I} \big( \big| t \big(Y(\mathbf{w}), \mathbf{w} \big) \big| \geq | t^{obs} | \big)P(\mathbf{W} = \mathbf{w}) \label{eqn:phiPValue}
\end{align}
where $\mathbb{W}^+_{\phi} = \{\mathbf{w}: \phi(\mathbf{w}, \mathbf{X}) = 1\}$ is the set of acceptable randomizations. The $p$-value $p_{\phi}$ is nearly identical to the $p$-value (\ref{randomizationTestPValue}), but using only the set of acceptable randomizations instead of the set of all randomizations. The set of acceptable randomizations is typically large, and thus the $p$-value $p_{\phi}$ cannot always be computed exactly. Instead, it can be unbiasedly estimated using
\begin{align}
	\hat{p}_{RS} &= \frac{ \sum_{m=1}^M \mathbb{I} \big( \big| t \big(Y(\mathbf{w}^{(m)}), \mathbf{w}^{(m)} \big) \big| \geq |t^{obs}| \big)}{M}, \text{ where } \mathbf{w}^{(m)} \sim P(\mathbf{W} | \phi(\mathbf{W}, \mathbf{X}) = 1) \label{eqn:rejectionSamplingPValue}
\end{align}
i.e., the approximation presented in (\ref{randomizationTestPValueApproximationSimple}). We propose a rejection-samping procedure for generating random samples $\mathbf{w}^{(1)},\dots,\mathbf{w}^{(M)} \sim P(\mathbf{W} | \phi(\mathbf{W}, \mathbf{X}) = 1)$: Randomly generate draws from $P(\mathbf{W})$, and only accept a draw $\mathbf{w}$ if $\phi(\mathbf{w}, \mathbf{X}) = 1$. For Bernoulli trials, this involves flipping $N$ coins (biased or unbiased, depending on the experimental design), and only accepting a particular assignment $\mathbf{w}$ if $\phi(\mathbf{w}, \mathbf{X}) = 1$.

While the rejection-sampling estimator $\hat{p}_{RS}$ is unbiased for $p_{\phi}$, it may be computationally intensive to generate random samples $\mathbf{w}^{(m)} \sim P(\mathbf{W} | \phi(\mathbf{W}, \mathbf{X}) = 1)$ if $\phi(\mathbf{W}, \mathbf{X})$ is particularly stringent. As an alternative, one can take an importance-sampling approach to biasedly estimate $p_{\phi}$ at a much lower computational cost.\cite{kong1992note,christian1999monte,robert2004monte} First, define a proposal distribution $P_q(\mathbf{W})$ whose support includes the support of $P(\mathbf{W} | \phi(\mathbf{W}, \mathbf{X}) = 1)$ but is less computationally burdensome to sample from than from $P(\mathbf{W} | \phi(\mathbf{W}, \mathbf{X}) = 1)$. Then, the importance-sampling estimator for $p_{\phi}$ is
\begin{align}
	\hat{p}_{IS} &= \frac{ \sum_{m=1}^M \mathbb{I} \big( \big| t \big(Y(\mathbf{w}^{(m)}), \mathbf{w}^{(m)} \big) \big| \geq |t^{obs}| \big) \frac{P(\mathbf{W} = \mathbf{w}^{(m)} | \phi(\mathbf{W}, \mathbf{X}) = 1)}{P_q(\mathbf{W} = \mathbf{w}^{(m)})} }{\sum_{m=1}^M \frac{P(\mathbf{W} = \mathbf{w}^{(m)} | \phi(\mathbf{W}, \mathbf{X}) = 1)}{P_q(\mathbf{W} = \mathbf{w}^{(m)})}}, \text{ where } \mathbf{w}^{(m)} \sim P_q(\mathbf{W})
\end{align}
In other words, the rejection-sampling estimator $\hat{p}_{RS}$ is a simple average based on the random draws $\mathbf{w}^{(m)} \sim P(\mathbf{W} | \phi(\mathbf{W}, \mathbf{X}) = 1)$, whereas the importance-sampling estimator is a weighted average based on the random draws $\mathbf{w}^{(m)} \sim P_q(\mathbf{W})$. Thus, $\hat{p}_{IS}$ will be easier to compute than $\hat{p}_{RS}$ if it is less computationally intensive to sample from the proposal distribution $P_q(\mathbf{W})$ than from the target distribution $P(\mathbf{W} | \phi(\mathbf{W}, \mathbf{X}) = 1)$.

The importance-sampling estimator can be reduced to a simple form by first noting that, under the assumption of a strongly ignorable assignment mechanism (\ref{eqn:psModel}),
\begin{align}
	P(\mathbf{W} = \mathbf{w} | \phi(\mathbf{W}, \mathbf{X}) = 1) &= \frac{P(\mathbf{W} = \mathbf{w}, \phi(\mathbf{W}, \mathbf{X}) = 1)}{P(\phi(\mathbf{W}, \mathbf{X}) = 1)} \\
	&= \frac{\prod_{i=1}^N e(\mathbf{x}_i)^{w_i} [1 - e(\mathbf{x}_i)]^{1 - w_i}}{P(\phi(\mathbf{W}, \mathbf{X}) = 1)}, \text{ where } \mathbf{w} \in \mathbb{W}^+_{\phi} \\
	&\propto \prod_{i=1}^N e(\mathbf{x}_i)^{w_i} [1 - e(\mathbf{x}_i)]^{1 - w_i}, \text{ where } \mathbf{w} \in \mathbb{W}^+_{\phi}
\end{align}
where $\mathbb{W}^+_{\phi} \equiv \{ \mathbf{w} \in \mathbb{W}^+ : \phi(\mathbf{w}, \mathbf{X}) = 1\}$ is the set of acceptable assignments according to the acceptance criterion. Then, if the proposal distribution is uniform across all acceptable assignments, i.e., if $P_q(\mathbf{W} = \mathbf{w}) = c$ for all $\mathbf{w} \in \mathbb{W}^+_{\phi}$, then the importance-sampling $p$-value approximation reduces to
\begin{align}
	\hat{p}_{IS} &= \frac{ \sum_{m=1}^M \mathbb{I} \big( \big| t \big(Y(\mathbf{w}^{(m)}), \mathbf{w}^{(m)} \big) \big| \geq |t^{obs}| \big) \prod_{i=1}^N e(\mathbf{x}_i)^{w^{(m)}_i} [1 - e(\mathbf{x}_i)]^{1 - w^{(m)}_i} }{\sum_{m=1}^M \prod_{i=1}^N e(\mathbf{x}_i)^{w^{(m)}_i} [1 - e(\mathbf{x}_i)]^{1 - w^{(m)}_i}}, \text{ where } \mathbf{w}^{(m)} \sim P_q(\mathbf{W}) \label{eqn:importanceSamplingPValue}
\end{align}
where the quantity $\prod_{i=1}^N e(\mathbf{x}_i)^{w^{(m)}_i} [1 - e(\mathbf{x}_i)]^{1 - w^{(m)}_i}$ is easy to compute because the propensity scores $e(\mathbf{x}_i)$ are known.

For example, sampling from the distribution $P(\mathbf{W} | \sum_{i=1}^N W_i = N_T)$ via rejection-sampling may be computationally intensive if the propensity scores differ across units and $N$ is large. One proposal distribution that is uniform across assignments is random permutations of $\mathbf{W}^{obs}$, whose support is equal to the support of $P(\mathbf{W} | \sum_{i=1}^N W_i = N_T)$ but is less computational to sample from. Thus, one can still utilize random permutations of $\mathbf{W}^{obs}$ to estimate the conditional randomization test $p$-value---as in Case 1 in Section \ref{ss:equalProbabilities}---using the importance-sampling estimator $\hat{p}_{IS}$.

However, as noted earlier, unlike the estimator $\hat{p}_{RS}$, the estimator $\hat{p}_{IS}$ is biased of order $M^{-1}$,\cite{kong1992note} which---as we show in Section \ref{s:simulationExample}---may break the validity of the conditional randomization test. Thus, we recommend using the rejection-sampling estimator $\hat{p}_{RS}$ to ensure valid inferences from our conditional randomization test if it is not computationally intensive to do so. However, if it is computationally intensive to generate draws $\mathbf{w} \sim P(\mathbf{W} | \phi(\mathbf{W}, \mathbf{X}) = 1)$ but easy to generate draws $\mathbf{w} \sim P_q(\mathbf{W})$ for some proposal distribution, then we recommend using the importance-sampling estimator $\hat{p}_{IS}$ while ensuring that the number of random samples $M$ is large such that the bias of $\hat{p}_{IS}$ is minimal. For an in-depth discussion of rejection-sampling versus importance-sampling, see Robert and Casella (Chapter 3).\cite{christian1999monte}

The above procedure is closely related to the rerandomization framework developed by Morgan and Rubin,\cite{morgan2012rerandomization} who define an assignment criterion $\phi(\mathbf{W}, \mathbf{X})$ in order to ensure a certain level of covariate balance as part of an experimental design. Recent works on rerandomization have shown how $\phi(\mathbf{W}, \mathbf{X})$ can be flexibly defined: Morgan and Rubin\cite{morgan2015rerandomization} defined $\phi(\mathbf{W}, \mathbf{X})$ such that it incorporates tiers of importance for covariates, and Branson et al.\cite{branson2016improving} defined $\phi(\mathbf{W}, \mathbf{X})$ such that it incorporates tiers of importance for both covariates and multiple treatment effects of interest.

However, the purpose of the introduction of $\phi(\mathbf{W}, \mathbf{X})$ here is to conduct a conditional randomization test, rather than yield a desirable experimental design. It is similar to the conditional randomization test of Hennessy et al.,\cite{hennessy2016conditional} who define $\phi(\mathbf{W}, \mathbf{X})$ in terms of categorical covariate balance. However, because Hennessy et al.\cite{hennessy2016conditional} and other conditional randomization tests (e.g., Rosenbaum\cite{rosenbaum1984conditional}) have focused on cases where propensity scores are equal across units or strata, they could sample from $P(\mathbf{W} | \phi(\mathbf{W}, \mathbf{X}) = 1)$ directly via random permutations of $\mathbf{W}^{obs}$. Indeed, both the rerandomization and conditional randomization test literature have focused on cases where the propensity scores are equal across units, whereas our approach addresses the more general case where propensity scores differ across units. Furthermore, if our rejection-sampling approach is computationally intensive, our importance-sampling approach allows one to still utilize random permutations of $\mathbf{W}^{obs}$ to quickly estimate the conditional randomization test $p$-value at the cost of incurring a small bias.

Now we establish that the unconditional and conditional randomization tests (i.e., the randomization test using $p$ in (\ref{randomizationTestPValue}) and the randomization test using $p_{\phi}$ in (\ref{eqn:phiPValue}), respectively) are valid tests for Bernoulli trial experiments. While these are results for the randomization tests that use the exact $p$-values $p$ and $p_{\phi}$, this also suggests that our rejection-sampling approach for unbiasedly estimating $p_{\phi}$ yields valid statistical inferences. In Section \ref{s:simulationExample}, we empirically confirm the validity of these randomization tests, and we discuss to what extent our importance-sampling approach also yields valid statistical inferences.

\subsection{Validity of Unconditional and Conditional Randomization Tests for Bernoulli Trial Experiments}

For both theorems presented below, we assume that the treatment is assigned according to the strongly ignorable assignment mechanism (\ref{eqn:psModel}). First, we establish that the randomization test that uses this assignment mechanism is valid, i.e., that the probability of this $\alpha$-level randomization test falsely rejecting the Sharp Null Hypothesis is no greater than $\alpha$. This result is unsurprising given well-known results about the validity of randomization tests. Then, we establish that the conditional randomization test---i.e., the randomization test that uses the assignment mechanism $P(\mathbf{W} | \phi(\mathbf{W}, \mathbf{X}) = 1)$ for some prespecified criterion $\phi(\mathbf{W}, \mathbf{X})$ instead of the assignment mechanism (\ref{eqn:psModel})---is also valid. This result is slightly surprising in the sense that the validity of the randomization test holds even if the test uses an assignment mechanism other than the one used to conduct the randomized experiment.

\begin{theorem}[Validity of Unconditional Randomization Test]
	\label{thm:unconditionalValidity}
	Assume that a randomized experiment is conducted using the strongly ignorable assignment mechanism (\ref{eqn:psModel}). Define the two-sided randomization-test $p$-value as
\begin{align}
    p \equiv \sum_{\mathbf{w} \in \mathbb{W}^+} \mathbb{I} \big( \big| t \big(Y(\mathbf{w}), \mathbf{w} \big) \big| \geq | t^{obs} | \big)P(\mathbf{W} = \mathbf{w}) \label{eqn:theorem1PValue}
\end{align}
for some test statistic $t \big(Y(\mathbf{W}), \mathbf{W} \big)$, where $\mathbb{W}^+ = \{0,1\}^N$. Then the randomization test that rejects the Sharp Null Hypothesis when $p \leq \alpha$ is a valid test in the sense that
\begin{align}
	P(p \leq \alpha | H_0) \leq \alpha
\end{align}
where $H_0$ is the Sharp Null Hypothesis defined in (\ref{sharpNull}). \\
\end{theorem}

\begin{theorem}[Validity of Conditional Randomization Test]
	\label{thm:conditionalValidity}
	Assume that a randomized experiment is conducted using the strongly ignorable assignment mechanism (\ref{eqn:psModel}). Define the two-sided conditional randomization-test $p$-value as
\begin{align}
    p_{\phi} \equiv \sum_{\mathbf{w} \in \mathbb{W}_{\phi}^+} \mathbb{I} \big( \big| t \big(Y(\mathbf{w}), \mathbf{w} \big) \big| \geq | t^{obs} | \big)P(\mathbf{W} = \mathbf{w}) \label{eqn:theorem2PValue}
\end{align}
for some test statistic $t \big(Y(\mathbf{W}), \mathbf{W} \big)$, where $\mathbb{W}_{\phi}^+ = \{\mathbf{w} \in \mathbb{W}^+: \phi(\mathbf{w}, \mathbf{X}) = 1\}$ is the set of acceptable randomizations according to some prespecified criterion $\phi(\mathbf{W}, \mathbf{X})$. Then the randomization test that rejects the Sharp Null Hypothesis when $p_{\phi} \leq \alpha$ is a valid test in the sense that
\begin{align}
	P(p_{\phi} \leq \alpha | H_0) \leq \alpha
\end{align}
where $H_0$ is the Sharp Null Hypothesis defined in (\ref{sharpNull}).	
\end{theorem}

The proofs for Theorems \ref{thm:unconditionalValidity} and \ref{thm:conditionalValidity} are in the Appendix. 

Now we illustrate our randomization test procedure using a simple example where the randomization test $p$-value is computed exactly. Then we conduct a simulation study where the randomization test $p$-value is estimated, and we compare the rejection-sampling and importance-sampling approaches for estimating the $p$-value. Furthermore, we empirically confirm the validity of our randomization tests as established by Theorems \ref{thm:unconditionalValidity} and \ref{thm:conditionalValidity} above, and we demonstrate how conditioning on various statistics of interest can be used to construct statistically powerful randomization tests for Bernoulli trial experiments.

\section{Simulation Study of Unconditional and Conditional Randomization Tests} \label{s:simulationExample}

\subsection{Illustrative Example: Computing the Exact $p$-value}

As discussed in Section \ref{ss:testingFishersSharpNull}, the randomization-test $p$-value is typically approximated using (\ref{randomizationTestPValueApproximationSimple}) by drawing many possible treatment assignments $\mathbf{w}^{(1)}, \dots, \mathbf{w}^{(M)}$. However, for small samples, the $p$-value can be computed exactly using (\ref{randomizationTestPValue}) by examining each $\mathbf{w}$ in the set of possible treatment assignments $\mathbb{W}^+$. Here we explore a small-sample example to illustrate how to conduct randomization tests and construct confidence intervals when propensity scores vary across units. We also discuss how this procedure differs from the typical case where propensity scores are the same across units.

Consider a randomized experiment with $N = 10$ units. The potential outcomes for these units are shown in Table \ref{tab:exampleN10}, where the true treatment effect is $\tau = 0.5$. Say that a randomized experiment characterized by Bernoulli trials has occurred; the corresponding propensity scores, treatment assignment, and observed outcomes are also shown in Table \ref{tab:exampleN10}. For now, assume that the task at hand is to conduct randomization-based inference for the average treatment effect given the treatment assignment, observed outcomes, and propensity scores in Table \ref{tab:exampleN10}.

\begin{table}[H]
\small\sf\centering
\begin{tabular}{cccccc}
\toprule
Unit $i$ & $Y_{i}$(0) & $Y_i$(1) & $W_i^{\text{obs}}$ & $y_{i}^{\text{obs}}$  & $e(\mathbf{x}_i)$\\
\midrule
1 & -0.56 & -0.06 & 0 & -0.56 & 0.1 \\
2 & -0.23 &  0.27 & 1 & 0.26 & 0.2 \\
3 &  1.56 &  2.06 & 1 & 2.06 & 0.3 \\
4 &  0.07 &  0.57 & 0 & 0.07 & 0.4 \\
5 &  0.13 &  0.63 & 0 & 0.13 & 0.5 \\
6 &  1.72 &  2.22 & 1 & 2.22 & 0.5 \\
7 &  0.46 &  0.96 & 1 & 0.96 & 0.6 \\
8 & -1.27 & -0.77 & 1 & -0.77 & 0.7 \\
9 & -0.69 & -0.19 & 0 & -0.69 & 0.8 \\
10& -0.45 &  0.05 & 1 & 0.05 & 0.9 \\
\bottomrule
\end{tabular}\\[2pt]
\caption{Potential outcomes, treatment assignment, observed outcome, and propensity score for 10 units in a hypothetical randomized experiment. Note that the true treatment effect is $\tau = 0.5$.}
\label{tab:exampleN10}
\end{table}

With $N = 10$ units, only $2^{10} = 1024$ possible treatment assignments can be considered. Excluding the treatment assignments $\mathbf{0}_N$ and $\mathbf{1}_N$ leaves 1022 possible assignments. Under the Sharp Null Hypothesis, the observed outcomes $\mathbf{y}^{obs}$ will be the same as those in Table \ref{tab:exampleN10} for all 1022 of these assignments. We test this hypothesis following the three-step procedure in Section \ref{ss:testingFishersSharpNull}: First choose $\mathbb{W}^+$ and $P(\mathbf{W})$, then choose a test statistic, and finally compute the randomization test $p$-value.

We first consider the set $\mathbb{W}^+ = \{0, 1\}^N \setminus (\mathbf{0}_N \cup \mathbf{1}_N )$ that was used during randomization, where
\begin{align}
     P(\mathbf{W} = \mathbf{w} | \mathbf{X}) = \frac{\prod_{i=1}^N e(\mathbf{x}_i)^{w_i}[1 - e(\mathbf{x}_i)]^{1 - w_i}}{1 - \prod_{i=1}^N e(\mathbf{x}_i) - \prod_{i=1}^N [1 - e(\mathbf{x}_i)]} 
 \end{align}
 for each $\mathbf{w} \in \mathbb{W}^+$, as previously shown in (\ref{eqn:biasedCoinRestricted01Probabilities}). We choose the mean-difference estimator---given in (\ref{eqn:meanDiffEstimator})---as the test statistic. We then iterate through each of the $1022$ treatment assignments $\mathbf{w} \in \mathbb{W}^+$ and compute the test statistic assuming the Sharp Null Hypothesis is true. Once this is done, the randomization test $p$-value can be computed exactly using
 \begin{align}
    P \big( |t \big(Y(\mathbf{W}), \mathbf{W} \big)| \geq |t^{obs}| \big) &= \sum_{\mathbf{w} \in \mathbb{W}^+} \mathbb{I} \big( \big| t \big(Y(\mathbf{w}), \mathbf{w} \big) \big| \geq | t^{obs} | \big)P(\mathbf{W} = \mathbf{w})
\end{align}
as previously shown in (\ref{randomizationTestPValue}). From Table \ref{tab:exampleN10}, one can calculate the observed test statistic, $t^{obs}$, which is equal to 1.06.

\begin{figure}[H]
\centering
\begin{subfigure}[t]{.55\textwidth}
\centering
\includegraphics[width=\linewidth]{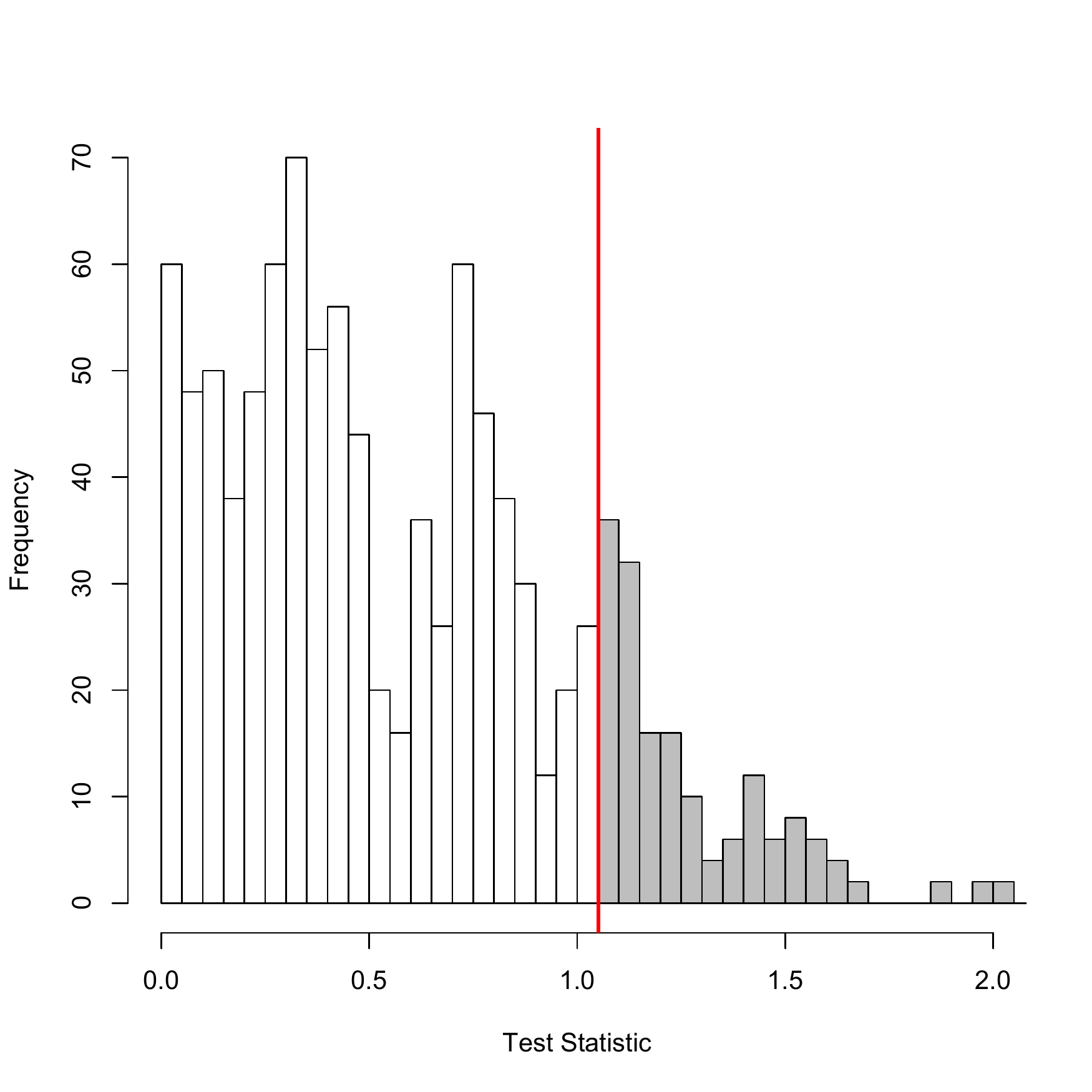}
\caption{The distribution of $t \big(Y(\mathbf{w}), \mathbf{w} \big)$ for each $\mathbf{w} \in \mathbb{W}^+$, where $\mathbb{W}^+ = \{0, 1\}^N \setminus (\mathbf{0}_N \cup \mathbf{1}_N )$. The observed test statistic is marked by a red vertical line. Assignments corresponding to test statistics larger than the observed one are in gray.  }
\label{fig:n10ExampleHistogram}
\end{subfigure}%

\begin{subfigure}[t]{.55\textwidth}
\centering
\includegraphics[width=\linewidth]{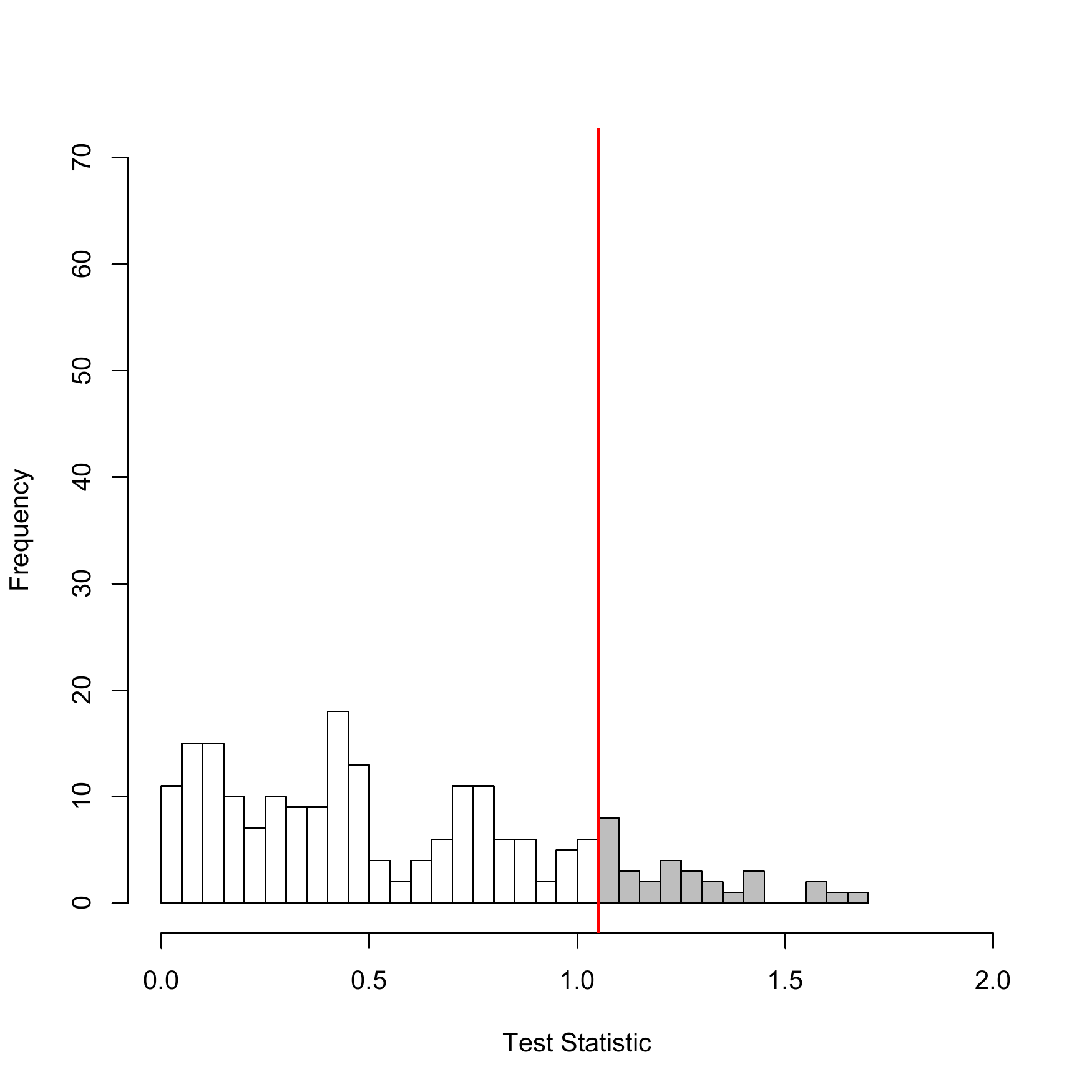}
\caption{The distribution of $t \big(Y(\mathbf{w}), \mathbf{w} \big)$ for each $\mathbf{w} \in \mathbb{W}^+$, where $\mathbb{W}^+ = \{ \mathbf{W} \in \mathbb{W} | \sum_{i=1}^N W_i = N_T \} $. }
\label{fig:n10ExampleConditionalHistogram}
\end{subfigure}

\caption{Unconditional and conditional randomization distributions of the test statistic under the Sharp Null Hypothesis.}

\end{figure}

Figure \ref{fig:n10ExampleHistogram} shows the distribution of the absolute value of the test statistic $t \big(Y(\mathbf{w}), \mathbf{w} \big)$ for each $\mathbf{w} \in \mathbb{W}^+$ assuming the Sharp Null Hypothesis is true. The portion of this distribution that corresponds to test statistics larger than the observed one is colored in gray. The randomization test $p$-value is then the probability of any gray treatment assignment occurring, which we find to be 0.12. If the propensity scores were equal across units---which is typically the case in the randomization test literature---then the randomization test $p$-value would simply be the number of gray treatment assignments divided by the total number of treatment assignments, which was, in this case, $\frac{164}{1022} \approx 0.16$. Thus, importantly, the $p$-value reflects the design of the randomized experiment---i.e., it incorporates the propensity scores that were used to randomize the units during the experiment.

Furthermore, we can obtain a confidence interval for the average treatment effect by inverting this randomization test using the procedure outlined in Section \ref{ss:confidenceIntervals}. We did a line search of values $\tau \in \{-3, -2.9, \dots, 2.9, 3\}$ and defined our 95\% confidence interval as the set of $\tau$'s for which we obtained $p$-values greater than 0.05 when testing the hypothesis (\ref{sharpNullTau}) for each $\tau$. We found the confidence interval to be $(-0.1, 2.4)$. Again, this confidence interval reflects the design of the randomized experiment, because the $p$-values corresponding to each $\tau$ depend on the propensity scores that were used during randomization.

Note that Figure \ref{fig:n10ExampleHistogram} displays every possible treatment assignment, including assignments where only one unit is assigned to treatment and the rest to control (and vice versa). However, researchers may want the statistical analysis to only consider treatment assignments similar to the observed one. For example, consider the more stringent set of treatment assignments $\mathbb{W}^+ = \{ \mathbf{W} \in \mathbb{W} | \sum_{i=1}^N W_i = N_T \}$, where in this example the number of treated units $N_T = 6$, as seen in Table \ref{tab:exampleN10}. Figure \ref{fig:n10ExampleConditionalHistogram} shows the distribution of the test statistic for each $\mathbf{w} \in \mathbb{W}^+$ in this case, assuming the Sharp Null Hypothesis is true. Note that there are only ${10 \choose 6} = 210$ treatment assignments, which is a subset of the assignments displayed in Figure \ref{fig:n10ExampleHistogram}. Again, the randomization test $p$-value is the probability of any gray treatment assignment occurring, but now the probability of any $\mathbf{w} \in \mathbb{W}^+$ is
\begin{align}
    P(\mathbf{W} = \mathbf{w} | \mathbf{X}) = \frac{\prod_{i=1}^N e(\mathbf{x}_i)^{w_i}[1 - e(\mathbf{x}_i)]^{1 - w_i}}{P(\sum_{i=1}^N W_i = N_T | \mathbf{X}) }
\end{align}
as previously shown in (\ref{eqn:probabilityConditionalNt}). Because there are only 210 treatment assignments $\mathbf{w}$ such that $\sum_{i=1}^N w_i = N_T$, we can compute the denominator exactly and thus compute the randomization test $p$-value exactly as well, which we find to be equal to 0.17. Furthermore, using the same procedure as above, we found the 95\% confidence interval to be $(-0.1, 2.4)$. Thus, in addition to reflecting the experimental design, randomization-based inference can also reflect particular experiments of interest, such as ones similar to the observed one.

Now we conduct a simulation study with $N = 100$ units. In this case, it is computationally intensive to compute randomization test $p$-values exactly, and we instead approximate them. Furthermore, because the propensity scores vary across units, it will be difficult to directly sample from conditional probability distributions such as $P(\mathbf{W} | \sum_{i=1}^N W_i = N_T)$, and thus we will need the rejection-sampling procedure from Section \ref{ss:acceptRejectProcedure} to conduct conditional inference.

\newpage

\thispagestyle{empty}

\newpage

\subsection{Simulation Setup}

Hennessy et al.\cite{hennessy2016conditional} conducted a simulation study to show that their randomization test that conditioned on categorical covariate balance was more powerful than unconditional randomization tests when covariates were associated with the outcome. Hennessy et al.\cite{hennessy2016conditional} consider the case where the propensity scores are the same across units. We modify their simulation study such that units' propensity scores differ. This simulation study serves two purposes:
\begin{enumerate}
    \item Confirm the validity of the unconditional and conditional randomization tests discussed in Section \ref{ss:unequalProbabilities}, as established by Theorems \ref{thm:unconditionalValidity} and \ref{thm:conditionalValidity}.
    \item Demonstrate how the rejection-sampling and importance-sampling procedures presented in Section \ref{ss:acceptRejectProcedure} can be used to construct statistically powerful conditional randomization tests.
\end{enumerate}
Consider $N = 100$ units with a single covariate $X$, where 50 units have covariate value $X = 1$ and the other 50 units have covariate value $X = 2$. Each unit has two potential outcomes---corresponding to treatment and control---which are generated once from the following:
\begin{equation}
\begin{aligned}
    Y_i(0) | X_i &\sim N(\lambda X_i, 1), \hspace{0.1 in} i = 1,\dots,N \\
    Y_i(1) &= Y_i(0) + \tau \label{eqn:potentialOutcomesModelSimulation}
\end{aligned}
\end{equation}
The parameter $\lambda$ determines the strength of the association between $X$ and the potential outcomes, while $\tau$ is the treatment effect. Similar to Hennessy et al.,\cite{hennessy2016conditional} we consider the values $\lambda \in \{0, 1.5, 3\}$ and $\tau \in \{0, 0.1, \dots, 1\}$ in our simulation. The previous example from Table \ref{tab:exampleN10} was generated using $\lambda = 0$ and $\tau = 0.5$.

The probability of the $i^{\text{th}}$ unit receiving treatment---i.e., its propensity score---was generated once from the following:
\begin{align}
    P(W_i = 1 | X_i) = P(W_i = 1) \sim \text{Beta}(5, 5), \hspace{0.1 in} i = 1,\dots, N \label{eqn:psModelSimulation}
\end{align}
This generating mechanism resulted in propensity scores being centered but spread around 0.5. In our simulation, propensity scores ranged from 0.22 to 0.87 with a mean of 0.49.

After the potential outcomes and propensity scores were generated, we randomly assigned units to treatment and control according to the probability distribution $P(\mathbf{W})$ defined by the propensity scores. We prevented any single treatment assignment from being $\mathbf{0}_N$ or $\mathbf{1}_N$; in other words, we considered the set of possible treatment assignments $\mathbb{W}^+ = \{0, 1\}^N \setminus (\mathbf{0}_N \cup \mathbf{1}_N )$ during randomization. In this case, there will always be 50 units with $X = 1$ and 50 units with $X = 2$, but the number of units assigned to treatment and control can vary from randomization to randomization. Any randomization of the 100 units to treatment and control can be summarized by Table \ref{tab:contingencyTable}, which includes the number of units assigned to treatment and control ($N_T$ and $N_C$) and the number of units with covariate values $X = 1$ and $X = 2$ ($N_1$ and $N_2$).

\begin{table}
\centering
    \begin{tabular}{ c c | c c | c }
    \hline
        & & \multicolumn{2}{ c |}{$\mathbf{W}$} &  \\
        & & 1 & 0 &  \\
        \hline
        \multirow{2}{*}{$X$} & 1 & $N_{T1}$ & $N_{C1}$ & $N_1 = 50$ \\
         & 2 & $N_{T2}$ & $N_{C2}$ & $N_2 = 50$ \\
         \hline
        & & $N_T$ & $N_C$ & $N = 100$ \\
        \hline
    \end{tabular}
    \caption{Contingency table of the number of units assigned to treatment and control ($N_T$ and $N_C$) and the number of units with covariate values $X = 1$ and $X = 2$ ($N_1$ and $N_2$). The values $N_1 = 50$, and $N_2 = 50$ were fixed across randomizations in the simulation study; the other values varied across randomizations.}
    \label{tab:contingencyTable}
\end{table}

Before conducting the full simulation, let's first consider one possible treatment assignment that we may observe during this simulation. We will present four randomization tests one could use to test the Sharp Null Hypothesis.

\subsection{Example of One Treatment Assignment} \label{ss:simulationExample}

Consider the case when $\lambda = 3$ and $\tau = 0.5$; i.e., when the covariate is strongly associated with the outcome and the treatment effect is moderate. The potential outcomes were generated using (\ref{eqn:potentialOutcomesModelSimulation}), the propensity scores were generated using (\ref{eqn:psModelSimulation}), and then units were randomized by flipping biased coins corresponding to these propensity scores. Table \ref{tab:exampleContingencyTable} shows the resulting randomization. Given this randomization and the corresponding dataset, how should we test the Sharp Null Hypothesis?

\begin{table}
\centering
    \begin{tabular}{ c c | c c | c }
    \hline
        & & \multicolumn{2}{ c |}{$\mathbf{W}^{obs}$} &  \\
        & & 1 & 0 &  \\
        \hline
        \multirow{2}{*}{$X$} & 1 & $N_{T1}^{obs} = 30$ & $N_{C1}^{obs} = 20$ & $N_1 = 50$ \\
         & 2 & $N_{T2}^{obs} = 24$ & $N_{C2}^{obs} = 26$ & $N_2 = 50$ \\
         \hline
        & & $N_T^{obs} = 54$ & $N_C^{obs} = 46$ & $N = 100$ \\
        \hline
    \end{tabular}
    \caption{Example of a possible treatment allocation in our simulation study.}
    \label{tab:exampleContingencyTable}
\end{table}

Any randomization test should involve generating treatment assignments via biased coins corresponding to the prespecified propensity scores, because this is how the randomization observed in Table \ref{tab:exampleContingencyTable} was generated. However, which set of possible treatment assignments $\mathbb{W}^+$ should one consider during the test? We consider four different $\mathbb{W}^+$ and their associated randomization tests:
\begin{enumerate}
    \item An unconditional randomization test (as presented in Section \ref{ss:testingFishersSharpNull}), with $\mathbb{W}^+ = \{0, 1\}^N \setminus (\mathbf{0}_N \cup \mathbf{1}_N )$.
    \item A randomization test conditional on the number of units assigned to treatment, with $\mathbb{W}^+ = \{ \mathbb{W} | \sum_{i=1}^N W_i = N_T^{obs} \}$.
    \item A randomization test conditional on the number of units with $X = 1$ assigned to treatment, with $\mathbb{W}^+ = \{ \mathbb{W} | \sum_{i: X_i = 1} W_i = N_{T1}^{obs} \}$.
    \item A randomization test conditional on $N_T$ and $N_{T1}$, with $\mathbb{W}^+ = \big\{ \mathbb{W} \big| \sum_{i=1}^N W_i = N_T^{obs} \hspace{0.05 in} \text{and }  \sum_{i: X_i = 1} W_i = N_{T1}^{obs} \big\}$.
\end{enumerate}
Arguably, the first randomization test is the most natural choice, because it corresponds to the $\mathbb{W}^+$ that was actually used to generate the randomization observed in Table \ref{tab:exampleContingencyTable}; however, because conditional randomization tests can be more powerful than unconditional randomization tests, the other three tests may be options researchers might consider as well.

The above tests are ordered in terms of the restrictiveness of $\mathbb{W}^+$: The first two randomization tests involve flipping biased coins to generate treatment assignments, where the values $N_{T1}$, $N_{C1}$, $N_{T2}$, and $N_{C2}$ in Table \ref{tab:contingencyTable} can vary across assignments; in the third randomization test, only $N_{T2}$ and $N_{C2}$ can vary; and in the fourth randomization test, none of these values can vary. Because iterating through every possible treatment assignment in $\mathbb{W}^+$ is computationally intensive---for the example in Table \ref{tab:exampleContingencyTable}, $|\mathbb{W}^+| = 2^{100} - 2$ for the first test, and $|\mathbb{W}^+| = {50 \choose 30}$ for the fourth test---we instead generate 1,000 treatment assignments $\mathbf{w}^{(1)}, \dots, \mathbf{w}^{(1000)}$ using our rejection-sampling procedure discussed in Section \ref{ss:acceptRejectProcedure} to approximate the randomization distribution for each test.

\begin{figure}[H]
\centering
    \includegraphics[scale = 0.5]{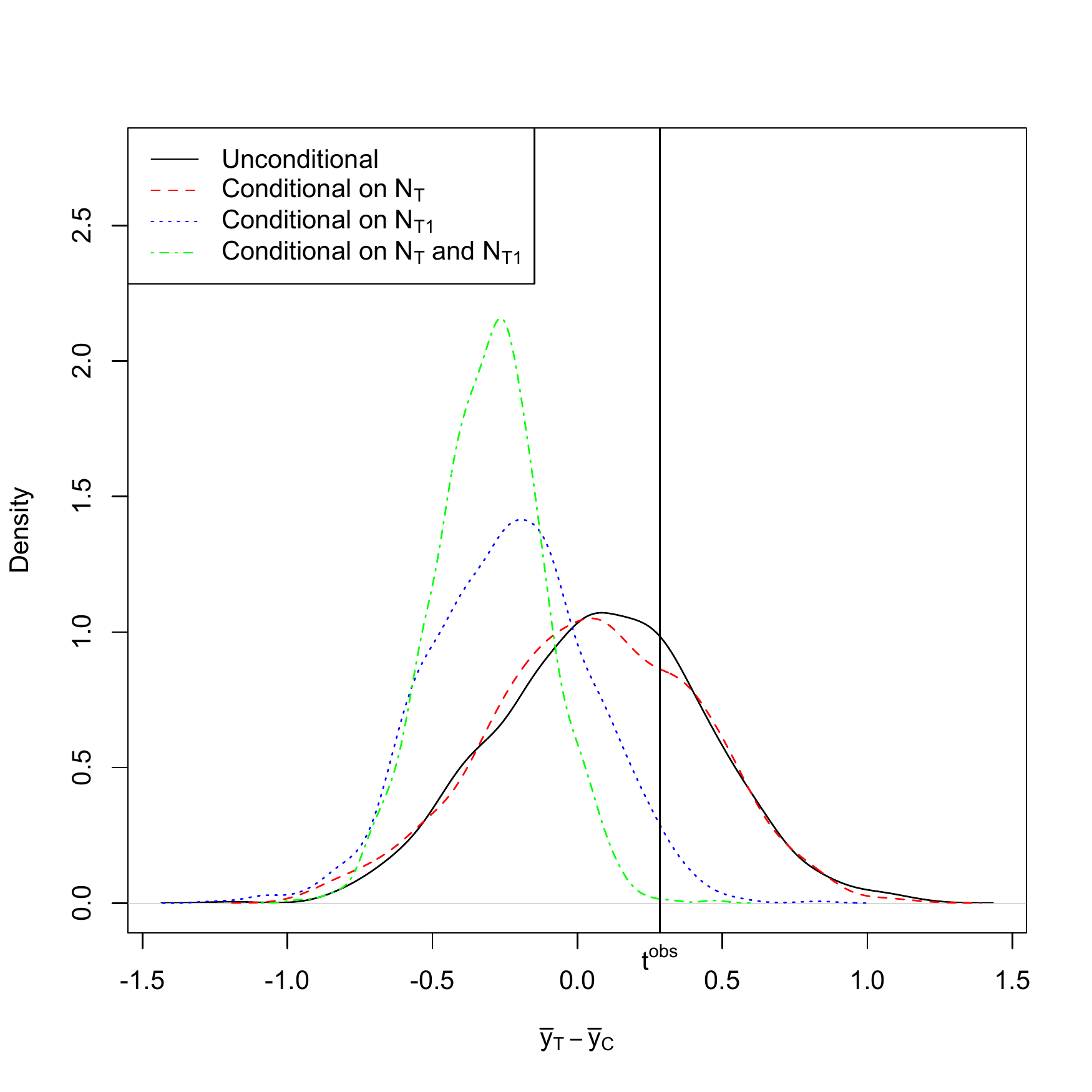}
    \caption{The unconditional and conditional randomization distributions for the mean-difference test statistic under the Sharp Null Hypothesis for the example in Table \ref{tab:exampleContingencyTable}. The observed test statistic for this example dataset is marked by a black vertical line. Each randomization distribution was approximated by drawing $\mathbf{w}^{(1)}, \dots, \mathbf{w}^{(1000)}$ from the corresponding $\mathbb{W}^+$ using the rejection-sampling procedure discussed in Section \ref{ss:acceptRejectProcedure}. }
    \label{fig:exampleRandomizationDistributions}
\end{figure}

The approximate randomization distribution of the mean-difference test statistic $\bar{y}_T - \bar{y}_C$ under the Sharp Null Hypothesis for each of these four tests is shown in Figure \ref{fig:exampleRandomizationDistributions}. The conditional randomization distributions for the third and fourth tests are shifted to the left of the unconditional randomization distribution. This is no coincidence: In Table \ref{tab:exampleContingencyTable}, there are more units with $X = 1$ in the treatment group and more units with $X = 2$ in the control group; as a result, the treatment group will have units with systematically lower potential outcomes, due to the potential outcomes model (\ref{eqn:potentialOutcomesModelSimulation}). This is reflected in the conditional randomization distributions but not the unconditional one. Consequentially, the conditional and unconditional randomization tests will give different results: One-sided $p$-values for the four tests are 0.58, 0.57, 0.08, and 0.00, respectively. This suggests that some of these randomization tests may be more powerful at detecting a treatment effect than others, which we further explore below.

\subsection{Full Simulation Study}

Now we compare the four randomization tests discussed in Section \ref{ss:simulationExample} in terms of their power. For each combination of $\lambda \in \{0, 1.5, 3\}$ and $\tau \in \{0, 0.1, \dots, 1\}$, the potential outcomes were generated using (\ref{eqn:potentialOutcomesModelSimulation}), the propensity scores were generated using (\ref{eqn:psModelSimulation}), and then units were randomized 1,000 times by flipping biased coins corresponding to these propensity scores.

\begin{figure}[H]
    \centering
    \includegraphics[scale = 0.75]{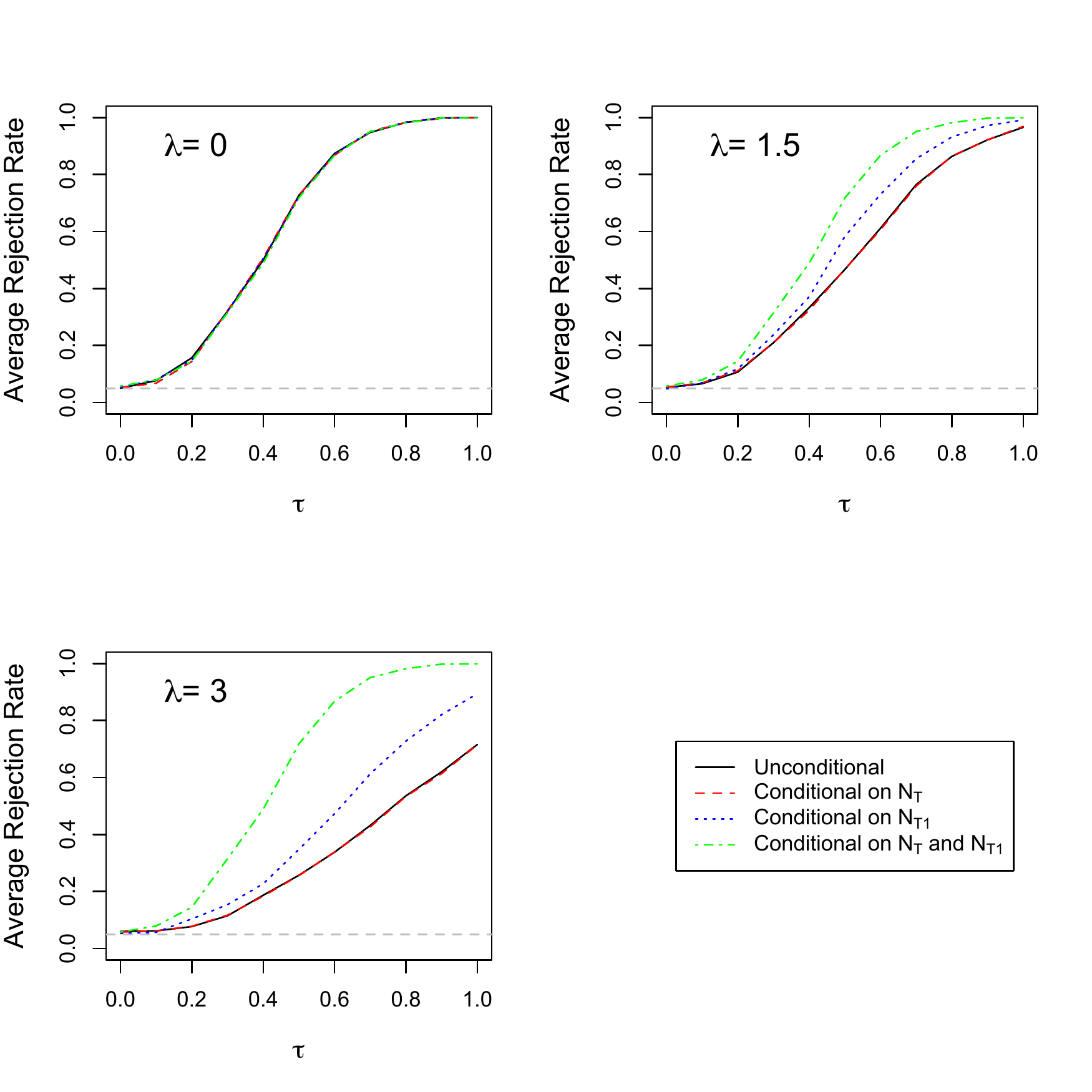}
    \caption{Average rejection rates for the four randomization tests across 1,000 randomizations for each value of $\lambda$ and $\tau$. As $\lambda$ increases, the covariate becomes more associated with the outcome; as $\tau$ increases, the treatment effect should become easier to detect. The gray horizontal line marks 0.05.}
    \label{fig:powerAnalysis}
\end{figure}

For each of the 1,000 randomizations, we performed the four randomization tests discussed in Section \ref{ss:simulationExample} using the rejection-sampling approach to unbiasedly estimate each $p$-value using $\hat{p}_{RS}$ given in (\ref{eqn:rejectionSamplingPValue}). For each test, we rejected the Sharp Null Hypothesis if $\hat{p}_{RS} \leq 0.05$. Figure \ref{fig:powerAnalysis} displays the average rejection rate of the Sharp Null Hypothesis---i.e., the power---for each randomization test. When $\tau = 0$, the Sharp Null Hypothesis is true, and all of the randomization tests reject the null 5\% of the time. This confirms the validity of our unconditional and conditional randomization tests, as established by Theorems \ref{thm:unconditionalValidity} and \ref{thm:conditionalValidity}. When $\lambda = 0$, the covariate is not associated with the outcome, and all of the randomization tests are essentially equivalent. As the covariate becomes more associated with the outcome, the third and fourth conditional randomization tests become more powerful than the unconditional test, while the randomization test that only conditions on $N_T$ remains equivalent to the unconditional randomization test. This is due to the fact that the quantity $N_{T1}$ combined with $N_T$ may be confounded with the treatment effect if there is covariate imbalance between the treatment and control groups, as in the example presented in Table \ref{tab:exampleContingencyTable} and Figure \ref{fig:exampleRandomizationDistributions}.

However, our rejection-sampling approach can be computationally expensive. Generating 1,000 samples for the unconditional randomization test, the randomization test conditional on $N_T$, the randomization test conditional on $N_{T1}$, and the randomization test conditional on $N_{T}$ and $N_{T1}$ took on average 0.25, 1.22, 2.14, and 34.75 seconds, respectively. As an alternative to the rejection-sampling approach for computing the randomization test $p$-value $\hat{p}_{RS}$ conditional on $N_T$ and $N_{T1}$, we can take our importance-sampling approach discussed in Section \ref{ss:acceptRejectProcedure}. Instead of sampling directly from $P(\mathbf{W} | \phi(\mathbf{W}, \mathbf{X}) = 1)$ via rejection-sampling, we generate $M$ proposals $\mathbf{w}^{(1)},\dots,\mathbf{w}^{(M)}$ uniformly from the set of acceptable randomizations $\{\mathbf{w}: \sum_{i=1}^N w_i = N_T \text{ and } \sum_{i: w_i = 1} \mathbb{I}(X_i = 1) = N_{T1}\}$; this corresponds to random permutations of $\mathbf{W}^{obs}$ within the $X = 1$ and $X = 2$ strata. Then, we compute $\hat{p}_{IS}$ given in (\ref{eqn:importanceSamplingPValue}) and reject if $\hat{p}_{IS} \leq 0.05$.

\begin{figure}[H]
    \centering
    \includegraphics[scale = 0.75]{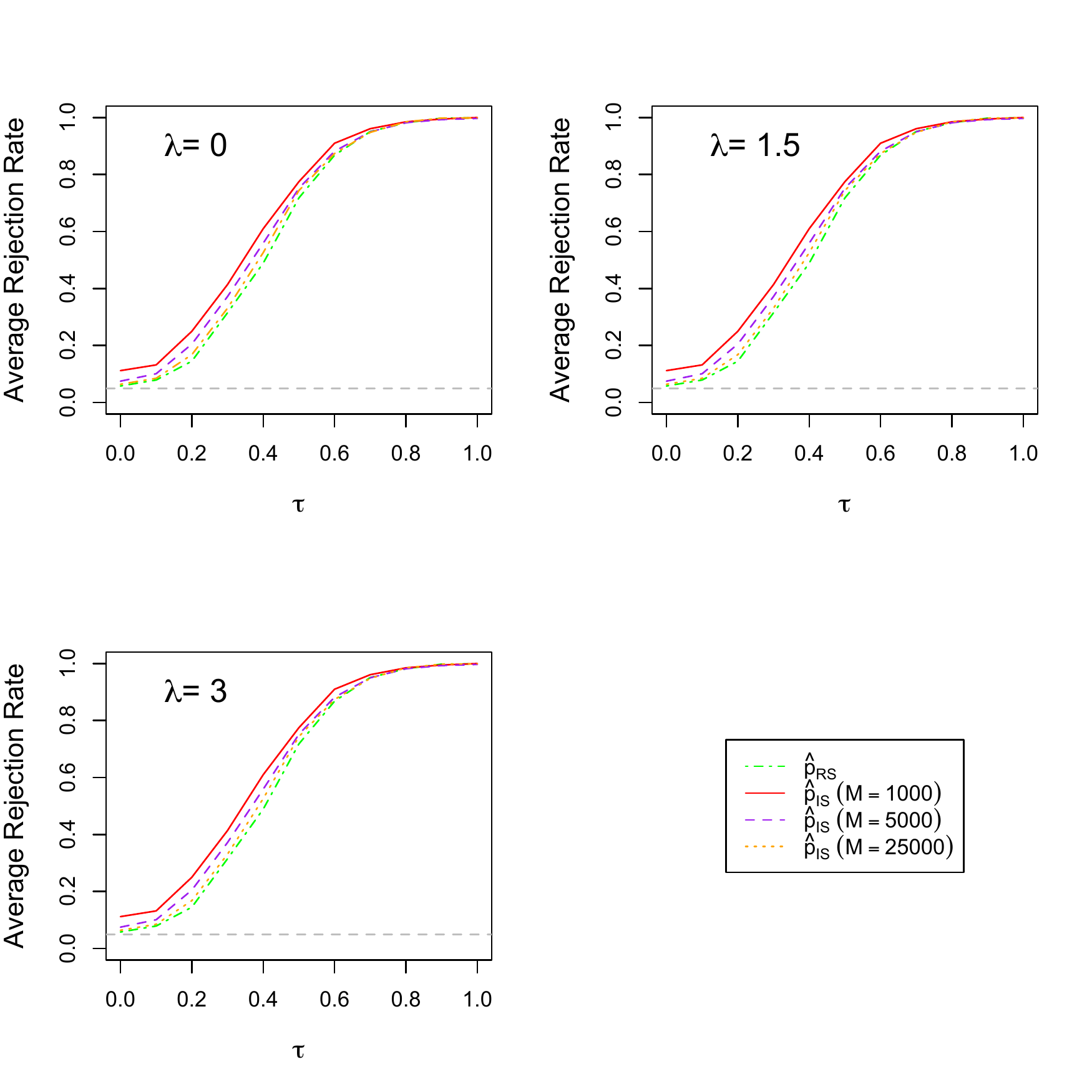}
    \caption{Average rejection rates for the rejection-sampling and importance-sampling approaches conditional on $N_T$ and $N_{T1}$. For importance-sampling, we tried various numbers of proposals $M$. The line for $\hat{p}_{RS}$ (i.e., the rejection-sampling approach) is the same as the line for ``Conditional on $N_T$ and $N_{T1}$'' in Figure \ref{fig:powerAnalysis}. }
    \label{fig:powerAnalysisRSvsIS}
\end{figure}

Figure \ref{fig:powerAnalysisRSvsIS} compares the rejection-sampling approach (i.e., rejecting the Sharp Null Hypothesis if $\hat{p}_{RS} \leq 0.05$) with the importance-sampling approach (i.e., rejecting the Sharp Null Hypothesis if $\hat{p}_{IS} \leq 0.05$) for different values of $M$. The importance-sampling approach is computationally less intensive than the rejection-sampling approach: The importance-sampling approach using $M = 1,000$, $M = 5,000$, and $M = 25,000$ took on average 0.68, 3.30, 16.31 seconds, respectively. Note that even the $M = 25,000$ case required less than half the time as the rejection-sampling approach. However, as noted in Section \ref{ss:acceptRejectProcedure}, $\hat{p}_{IS}$ has a bias of order $M^{-1}$, and thus the $p$-value for the importance-sampling approach may be notably biased for low $M$. This can be seen in Figure \ref{fig:powerAnalysisRSvsIS}: For $M = 1,000$, the importance-sampling approach falsely rejects the Sharp Null Hypothesis when $\tau = 0$ at a substantially higher rate than 0.05; this suggests that the importance-sampling approach has a negative bias in this case. However, as $M$ increases, this bias is less substantial, and results using $\hat{p}_{IS}$ approach those using $\hat{p}_{RS}$. Thus, the bias of importance-sampling can break the validity of our randomization test, but this can be alleviated by increasing the number of proposals $M$ at a minimal computational cost.

In summary, these results reinforce the idea of Hennessy et al.\cite{hennessy2016conditional} that conditional randomization tests are more powerful than unconditional randomization tests when the acceptance criterion $\phi(\mathbf{W}, \mathbf{X})$ incorporates statistics that are associated with the outcome. Furthermore, this demonstrates how our rejection-sampling procedure can be used to condition on several combinations of statistics of interest, thus yielding statistically powerful randomization tests for Bernoulli trial experiments. Finally, when this rejection-sampling procedure is computationally intensive, our importance-sampling approach is a viable alternative; however, we recommend generating a large number of proposals $M$ such that the bias of the importance-sampling approach is neglible and thus still yields valid statistical inferences.

\section{Discussion and Conclusion} \label{s:discussion}

Here we presented a randomization-based inference framework for experiments whose assignment mechanism is characterized by independent Bernoulli trials. Our framework and corresponding randomization tests encapsulate all strongly ignorable assignment mechanisms, including experiments based on complete, blocked, and paired randomization, as well as the general case where propensity scores differ across all units. In particular, we introduced rejection-sampling and importance-sampling approaches for obtaining randomization-based point estimates and confidence intervals conditional on any statistics of interest for Bernoulli trial experiments, which has not been previously studied in the literature. We also established that our randomization test is a valid test, and the power of this test can be improved by conditioning on various statistics of interest without sacrificing the validity of the test.

While our discussion of point estimates and confidence intervals are based on a sharp hypothesis that assumes a constant additive treatment effect, our framework can be extended to any sharp hypothesis, including those that incorporate heterogeneous treatment effects. Recent works in the randomization-based inference literature have begun to address treatment effect heterogeneity (e.g., Ding et al.\cite{ding2015randomization} and Caughey et al.\cite{caughey2016beyond}), and our framework can be extended to these discussions.

Throughout, we assumed that the propensity scores are known, as in randomized experiments.  In observational studies, the propensity scores are estimated, typically with model-based methodologies like logistic regression. Nonetheless, propensity score methodologies still assume a strongly ignorable assignment mechanism as in (\ref{eqn:psModel}), with the assumption that the estimated propensity scores $\hat{e}(\mathbf{x})$ are ``close'' to the true $e(\mathbf{x})$, i.e., the propensity score model is well-specified. An implication of our randomization-based inference framework is that it can still be applied to observational studies, where estimates $\hat{e}(\mathbf{x})$ are used instead of known $e(\mathbf{x})$. Such a test is valid to the extent that the $\hat{e}(\mathbf{x})$ are ``close'' to the true $e(\mathbf{x})$; this is not a limitation of our framework specifically but of propensity score methodologies in general. Determining when our randomization test is valid for observational studies is future work.

However, our randomization test would seem to be the most natural randomization test to use for observational studies, because it directly reflects the strongly ignorable assignment mechanism (\ref{eqn:psModel}) that is assumed in most of the observational study literature. Other proposed randomization tests for observational studies reflect other assignment mechanisms, such as blocked and paired assignment mechanisms; these randomization tests are not immediately applicable to cases where the propensity score varies across all units.

There are many other methodologies for analyzing randomized experiments and observational studies, such as regression with or without inverse probability weighting, matching, and Bayesian modeling. Importantly, all of these methodologies assume the strongly ignorable assignment mechanism (\ref{eqn:psModel}) in addition to other assumptions about model specification, asymptotics, or units' propensity scores within covariate strata. Our framework only makes the strongly ignorable assignment mechanism assumption, and thus is a minimal-assumption approach while still yielding point estimates and confidence intervals that directly reflect the assignment mechanism. Furthermore, we established the validity of our randomization test and demonstrated how conditioning on relevant statistics of interest can yield powerful randomization tests for Bernoulli trial experiments.

\newpage

\section{Appendix}

\subsection{Proof of Theorem \ref{thm:unconditionalValidity}}

This proof closely follows the proof provided in Hennessey et al. (Page 64),\cite{hennessy2016conditional} but with a focus on Bernoulli trial experiments instead of completely randomized experiments.

Define $T_W$ as a random variable whose distribution is the same as $|t(Y(\mathbf{W}), \mathbf{W})|$, for some test statistic $t(Y(\mathbf{W}), \mathbf{W})$, where $\mathbf{W} \sim P(\mathbf{W} | \mathbf{X})$ specified by the strongly ignorable assignment mechanism (\ref{eqn:psModel}). Furthermore, let $F_{T_W}(\cdot)$ be the CDF of $T_W$. Note that $T_W$ must be defined for all $\mathbf{W} \in \mathbb{W}^+$, including $\mathbf{W} = \mathbf{1}_N$ or $\mathbf{W} = \mathbf{0}_N$; without loss of generality, one can define $T_W = 0$ for these two cases. 

Under the Sharp Null Hypothesis $H_0$ defined in (\ref{sharpNull}), $Y(\mathbf{W}) = \mathbf{y}^{obs}$ for all $\mathbf{W} \in \mathbb{W}^+$. Thus, under $H_0$,
\begin{align}
	|t(\mathbf{y}^{obs}, \mathbf{W})| \sim T_W \label{eqn:testStatisticNullDistribution}
\end{align}
i.e., the distribution of the observed test statistic $|t^{obs}| \equiv |t(\mathbf{y}^{obs}, \mathbf{W}^{obs})|$ across randomizations is the same as the distribution of $T_W$.

Now note that the randomization test $p$-value defined in (\ref{eqn:theorem1PValue}) of Theorem \ref{thm:unconditionalValidity} is such that, under $H_0$,
\begin{align}
	p &= 1 - F_{T_W}(|t^{obs}|)
\end{align}
Furthermore, given (\ref{eqn:testStatisticNullDistribution}), we have that the distribution of $p$ across randomizations is
\begin{align}
	p &\sim 1 - F_{T_W}(T_W)
\end{align}
under $H_0$.

If $T_W$ were continuous, then $(1 - F_{T_W}(T_W)) \sim \text{Unif}(0,1)$ by the probability integral transform; however, $T_W$ is discrete due to the discreteness of $\mathbb{W}^+$. Nonetheless, $(1 - T_W)$ stochastically dominates $U \sim \text{Unif}(0,1)$, and thus
\begin{align}
	P(p \leq \alpha | H_0) &\leq P(U \leq \alpha | H_0) \label{eqn:stochasticDominance} \\
	&\leq \alpha \label{eqn:uniformDistribution}
\end{align}
where (\ref{eqn:stochasticDominance}) follows from the definition of stochastic dominance, and (\ref{eqn:uniformDistribution}) follows from properties of the standard uniform distribution. This concludes the proof of Theorem \ref{thm:unconditionalValidity}.

\subsection{Proof of Theorem \ref{thm:conditionalValidity}}

Define a set of partitions $\mathbb{W}^+_1,\dots,\mathbb{W}^+_B$, where $\mathbb{W}^+_b \cap \mathbb{W}^+_{b'} = \emptyset$ for all $b \neq b'$ and $\cup_{b=1}^B \mathbb{W}^+_b = \mathbb{W}^+ = \{0,1\}^N$. In other words, the $\mathbb{W}^+_1,\dots,\mathbb{W}^+_B$ partition the set of possible randomizations under the strongly ignorable assignment mechanism (\ref{eqn:psModel}) into non-overlapping sets. Consider a randomization test that is conducted only within a particular one of these partitions; the associated randomization test $p$-value is
\begin{align}
    p_b \equiv \sum_{\mathbf{w} \in \mathbb{W}_b^+} \mathbb{I} \big( \big| t \big(Y(\mathbf{w}), \mathbf{w} \big) \big| \geq | t^{obs} | \big)P(\mathbf{W} = \mathbf{w})
\end{align}
Importantly, by Theorem \ref{thm:unconditionalValidity}, for randomizations $\mathbf{W} \in \mathbb{W}^+_b$, the randomization test that rejects the Sharp Null Hypothesis when $p_b \leq \alpha$ is a valid test, i.e.,
\begin{align}
	P(p_b \leq \alpha | H_0, \mathbf{W} \in \mathbb{W}^+_b) \leq \alpha \text{ for all } b = 1,\dots,B
\end{align}

The acceptance criterion $\phi(\mathbf{W}, \mathbf{X})$ determines the particular partition in which the conditional randomization test is conducted. Without loss of generality, say that $\phi(\mathbf{W}, \mathbf{X})$ is defined such that
\begin{align}
	\phi(\mathbf{W}, \mathbf{X}) \equiv \begin{cases}
		1 &\mbox{ if } \mathbf{W} \in \mathbb{W}^+_b \text{ for some } b =1,\dots,B \\
		0 &\mbox{ otherwise.}
	\end{cases}	
\end{align}

Defined this way, $\phi(\mathbf{W}, \mathbf{X})$ varies across randomizations $\mathbf{W} \in \mathbb{W}^+$; as a result, the set of acceptable randomizations $\mathbb{W}^+_{\phi} \equiv \{ \mathbf{w} \in \mathbb{W}^+: \phi(\mathbf{w}, \mathbf{X}) = 1 \}$ varies across $\mathbf{W} \in \mathbb{W}^+$ as well. As an example, consider the criterion $\phi(\mathbf{W}, \mathbf{X})$ defined as equal to $1$ if $\sum_{i=1}^N W_i = N_T$ and equal to $0$ otherwise. The number of treated units $N_T$ can vary across $\mathbf{W} \in \mathbb{W}^+$, and thus $\mathbb{W}^+_{\phi}$ will vary across $\mathbf{W} \in \mathbb{W}^+$ as well, based on the realization of $N_T$. In this case, the partitions $\mathbb{W}^+_1,\dots,\mathbb{W}^+_B$ are defined as the sets of treatment assignments corresponding to the unique values of $N_T$. In general, the criterion $\phi(\mathbf{W}, \mathbf{X})$ will be a function of statistics, and the partitions $\mathbb{W}^+_1,\dots,\mathbb{W}^+_B$ can be defined by the unique values of these statistics. This setup is a generalization of the covariate balance function discussed in Hennessy et al. (Page 67).\cite{hennessy2016conditional}

Thus, for each $b=1,\dots,B$, there is an associated probability $P(\mathbf{W} \in \mathbb{W}^+_b) = P(\mathbb{W}^+_{\phi} = \mathbb{W}^+_b)$, and this probability is determined by the strongly ignorable assignment mechanism (\ref{eqn:psModel}). Once it is determined which partition the set of acceptable randomizations is equal to, the randomization test is conducted within this partition; i.e., the $p$-value $p_b$ is used for the $b$ such that $\mathbb{W}^+_{\phi} = \mathbb{W}^+_b$.

Thus, for the conditional randomization test $p$-value $p_{\phi}$ defined in Theorem \ref{thm:conditionalValidity}, we have that
\begin{align}
	P(p_{\phi} \leq \alpha | H_0) &= \sum_{b=1}^B P(p_{\phi} \leq \alpha | H_0, \mathbb{W}^+_{\phi} = \mathbb{W}^+_b) P(\mathbb{W}^+_{\phi} = \mathbb{W}^+_b) \text{ (by law of total probability)} \\
	&= \sum_{b=1}^B P(p_b \leq \alpha | H_0, \mathbb{W}^+_{\phi} = \mathbb{W}^+_b) P(\mathbb{W}^+_{\phi} = \mathbb{W}^+_b) \\
	&\leq \sum_{b=1}^B \alpha P(\mathbb{W}^+_{\phi} = \mathbb{W}^+_b) \text{ (by Theorem \ref{thm:unconditionalValidity})} \\
	&= \alpha \text{ (because $\sum_{b=1}^B P(\mathbb{W}^+_{\phi} = \mathbb{W}^+_b) = 1$)}
\end{align}
which is our desired result.

\newpage

\bibliography{bernoulliTrialsBib}

\bibliographystyle{vancouver}

\end{document}